\documentclass[%
 aip,
 pof,
 amsmath,
 amssymb,
 footinbib,
 twocolumn,
 floatfix,
 reprint
]{revtex4-1}

\usepackage[utf8]{inputenc}
\usepackage{CJKutf8}

\usepackage{mathrsfs}
\usepackage{bm}

\usepackage{graphicx}
\usepackage[subrefformat=parens,labelformat=parens,caption=false]{subfig} 

\usepackage{dcolumn} 
\usepackage{color}
\usepackage{siunitx}
\usepackage{faktor}

\usepackage{hyperref}
\hypersetup{
    colorlinks=true,
    breaklinks=true
}

\newcommand{\rd}{\mathrm{d}}

\allowdisplaybreaks

\begin{document}

\begin{CJK*}{UTF8}{} 

\title{Reduced models of unidirectional flows in compliant rectangular ducts at finite Reynolds number}
\thanks{This paper is based in part on a contributed paper\citep{WC20} published as part of the Proceedings of the 12th International On-line Conference for Promoting the Application of Mathematics in Technical and Natural  Sciences -- AMiTaNS’20.}

\author{Xiaojia Wang (\CJKfamily{gbsn}汪小佳)}
 \email{wang4142@purdue.edu}
\author{Ivan C.\ Christov}
 \email[Corresponding author:~]{christov@purdue.edu}
 \homepage{http://tmnt-lab.org}
\affiliation{School of Mechanical Engineering, Purdue University, West Lafayette, Indiana 47907, USA}

\begin{abstract}
Soft hydraulics, which addresses the interaction between an internal flow and a compliant conduit, is a central problem in microfluidics. We analyze Newtonian fluid flow in a rectangular duct with a soft top wall at steady state. The resulting fluid--structure interaction (FSI) is formulated for both vanishing and finite flow inertia. At the leading-order in the small aspect ratio, the lubrication approximation implies that the pressure only varies in the streamwise direction. Meanwhile, the compliant wall's slenderness makes the fluid--solid interface behave like a Winkler foundation, with the displacement fully determined by the local pressure. Coupling flow and deformation and averaging across the cross-section leads to a one-dimensional reduced model. In the case of vanishing flow inertia, an effective deformed channel height is defined rigorously to eliminate the spanwise dependence of the deformation. It is shown that a previously-used averaged height concept is an acceptable approximation. From the one-dimensional model, a friction factor and the corresponding Poiseuille number are derived. Unlike the rigid duct case, the Poiseuille number for a compliant duct is not constant but varies in the streamwise direction. Compliance can increase the Poiseuille number by a factor of up to four. The model for finite flow inertia is obtained by assuming a parabolic vertical variation of the streamwise velocity. To satisfy the displacement constraints along the edges of the channel, weak tension is introduced in the streamwise direction to regularize the Winkler-foundation-like model. Matched asymptotic solutions of the regularized model are derived.
\end{abstract}

\maketitle

\end{CJK*}

\section{Introduction}

As Frank M.\ White notes in \S3-3.3 of his iconic \textit{Viscous Fluid Flow} textbook\cite{W06_book}, ``fully developed duct flow is equivalent to a classic Dirichlet problem, [thus] it is not surprising that an enormous number of exact solutions are known.'' He (and, also, \citet{B08}) summarize an elegant set of solutions for such unidirectional flows, which are \emph{exact} solutions of the incompressible Navier--Stokes equations at \emph{any} Reynolds number, including ``lima\c{c}on-shaped ducts, for example, [which] are not commercially available at present'' \cite{W06_book}, adding a bit of humour to this topic. The general result is that, for all such duct flows, the volumetric flow rate $q$ is related to the axial pressure gradient, $-\rd p/\rd z$, via some (likely complicated) function of the cross-sectional geometry. This result is the cornerstone of \emph{hydraulics}, \textit{i.e.}, ``the conveyance of liquids through pipes and channels'' (per Oxford Languages---the provider of Google's English dictionary), a topic that is taught to undergraduate students \cite{P11}.

Now, however, what if the duct were manufactured from a soft material so that the local hydrodynamic pressure changes the cross-sectional area? Such problems have a time-honored history in biomechanics \cite{RK72,P80,GJ04} but not so much in hydraulics. Nevertheless, with the emergence of \emph{microfluidics} \cite{SSA04,SQ05,W06,B08,C13_book}, the hydraulics of compliant ducts manufactured from polymeric materials has become a central problem at the intersection of fluid mechanics and soft matter physics. To develop a theory of \emph{soft hydraulics}, we must understand steady fluid--structure interactions (FSIs). FSIs between external or internal flows (either viscous or inviscid) and elastic structures, as well as the linear stability of such coupled mechanics problems, is also a well-developed research subject  \cite{P16}, including fast progress in the last decade \cite{KCC18}. While FSI topics such as aeroelasticity \cite{BAH96} and moderate-Reynolds-number blood flow in large arteries \cite{P80} are now quite classical, the mechanical interaction between \emph{slow} viscous flows and \emph{compliant} conduits \cite{CPFY12} has opened new avenues of FSI research \cite{DS16,KCC18}, both at the microscale for, \textit{e.g.}, lab-on-a-chip applications \cite{CPFY12,FZPN19}, and at the macroscale for, \textit{e.g.}, soft robotics applications \cite{MEG17,Polygerinos17}.

To this end, in this paper, the soft hydraulics and its mathematical formulation are first introduced in section~\ref{sec:problem_statement}. Then, the discussion bifurcates into the case of a vanishing Reynolds number (section~\ref{sec:zero-Re}) and the case of a finite Reynolds number (section~\ref{sec:nonzero-Re}). We review the key recent results regarding flows in compliant ducts of initially rectangular cross-section. Then, within each of sections~\ref{sec:zero-Re} and \ref{sec:nonzero-Re}, we show how to consistently reduce these inherently three-dimensional (3D) problems to two-dimensional (2D) problems \footnote{Note that these are 2D problems in an axial vertical $(y,z)$ plane, using the axes notation in figure~\ref{fig:schematic},  not an $(x,z)$ plane perpendicular to the flow direction as in White's 2D Dirichlet problems for unidirectional duct flows \cite{W06_book}.} and, eventually, to one-dimensional (1D) models (that only involve axial, or streamwise, variations)\footnote{Since the issue of ``dimensionality'' of fluid flows and models has caused some confusion in the literature, here we restate, from \citet{P11}, the accepted definition that we shall employ: ``[a] flow is classified as one-, two-, or three-dimensional depending on the number of space coordinates required to specify the velocity field'' (p.~24).}. { Specifically, in section~\ref{sec:zero-Re}, we compare our consistent formulation with previous spanwise-averaged (\textit{i.e.}, over $x$, see figure~\ref{fig:schematic}) models, and ascertain the accuracy of the previous approach to the hydraulic predictions.}  Towards this end, in section~\ref{sec:friction_factor}, we introduce a generalization of the laminar flow friction factor suitable for quantifying the effect of compliance in soft hydraulic systems. In section~\ref{sec:nonzero-Re}, we address the issue of Reynolds number dependence (flow acceleration), which is a novel contribution of our work to the field of soft hydraulics. However, the model breaks down beyond a certain Reynolds number, requiring a regularization (section \ref{sec:R1_reg}), which leads to an interesting singular perturbation problem (solved in appendix~\ref{sec:app}). Finally, conclusions and avenues for future work are discussed in section~\ref{sec:conclusion}.


\section{Preliminaries, notation, and problem statement}
\label{sec:problem_statement}

\begin{figure}
    \centering
    \includegraphics[trim=1.5cm 0cm 2cm 0cm, clip, width=\columnwidth]{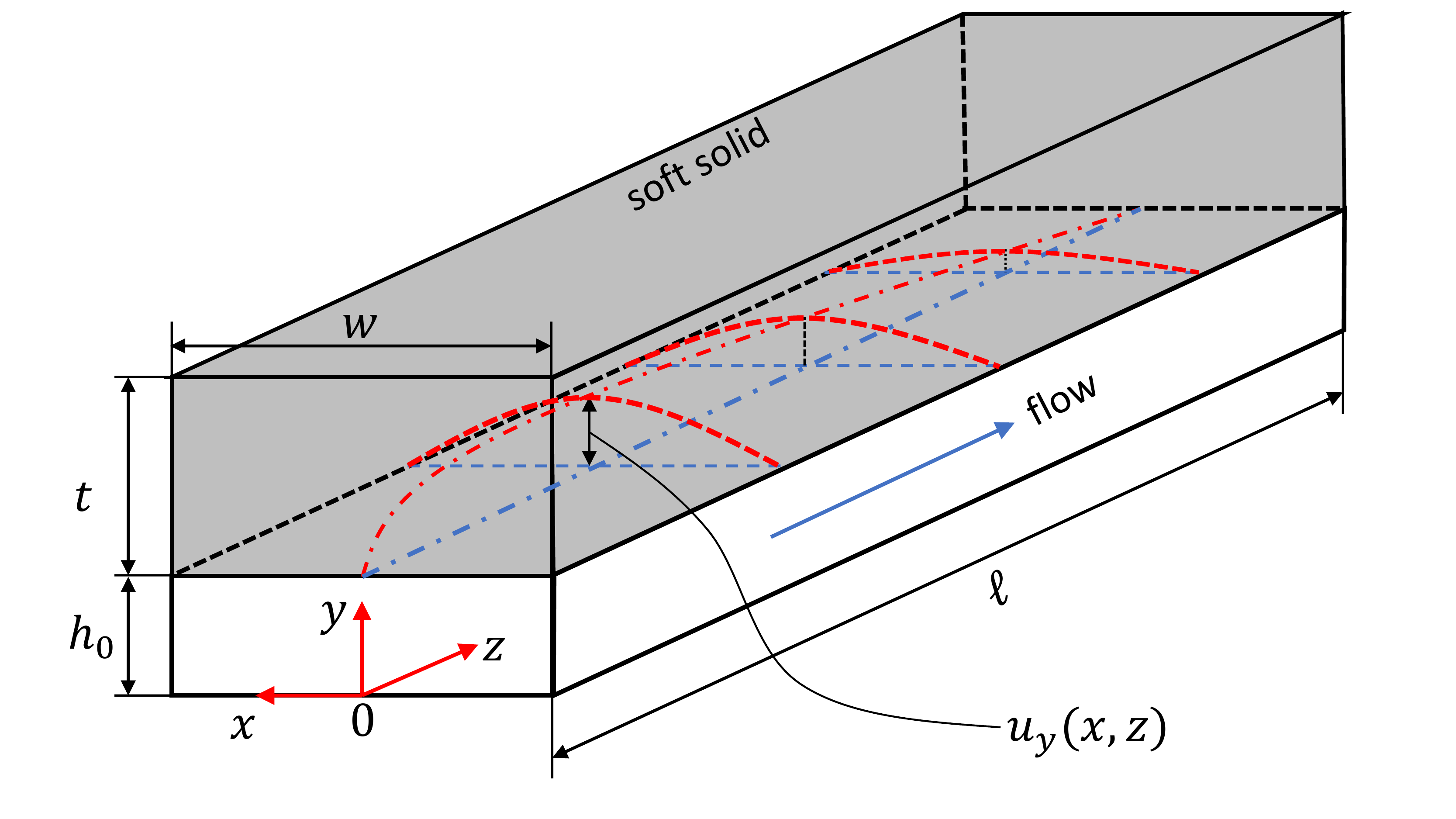}
    \caption{Diagram of a long, shallow rectangular microchannel with a compliant top wall, labelled with the dimensional variables of the problem (denoted by lower case letters and symbols). The origin of the coordinate system is set at the centerline ($x=0$) of the rigid bottom wall of the channel ($y=0$). The deformed fluid--solid interface is defined as $y=h_0+u_y(x,z)$, where $u_y$ denotes the compliant top wall's $y$-displacement evaluated at $y=h_0$. The Newtonian fluid flow, with a given volumetric flow rate $q$, is in the positive $z$-direction, as indicated by arrows, from the inlet at $z=0$ to the outlet at $z=\ell$. Exemplar deformation profiles of the fluid--solid interface at different streamwise locations are shown by the red dashed curves, while the interface deformation along $x=0$ is represented by the red dash-dotted curve. Zero displacement conditions are enforced along $z=0$, $z=\ell$, $x=-w/2$, and $x=w/2$.}
    \label{fig:schematic}
\end{figure}

Consider a soft-walled microchannel (initially a rectangular duct), which exhibits flow-induced deformation due to Newtonian fluid flow through it \cite{GEGJ06}. Denote the channel's undeformed height, width, length and top wall thickness by $h_0$, $w$, $\ell$, and $t$, respectively, as in figure~\ref{fig:schematic}. Further, introducing the aspect ratios, $\delta = h_0/w$ and $\epsilon=h_0/\ell$, we say that the microchannel is long and shallow  \cite{CCSS17} if $\epsilon\ll\delta\ll 1$. This kind of compliant duct is a common outcome of rapid microfluidic device fabrication via soft lithography \cite{XW98,SMMC11}. In the following analysis, the top wall's deformation is dominant, and thus it is the only deformation of interest \cite{CCSS17,WC19}. Denote the deformed cross-sectional height by $h(x,z)$ and assume the smallness of the aspect ratios still holds in the deformed microchannel, \textit{i.e.}, the deformed channel height is such that $h(x,z)\ll w\ll \ell$. { Note that $h(x,z)=h_0 + u_y(x,z)$ is the deformed channel height, where $u_y(x,z)$ is the displacement of the fluid--solid interface.} 

The incompressible Navier--Stokes (iNS) equations at steady state govern the flow within the duct. The velocity field is denoted $\bm{v} = (v_x,v_y,v_z)$ in Cartesian coordinates. { To make iNS dimensionless, let us introduce the following dimensionless variables \cite{CCSS17, WC19} (denoted by capital letters):
\begin{multline}\label{var_dim}
    X=\frac{x}{w},\quad Y=\frac{y}{h_0}, \quad Z=\frac{z}{\ell}, \\
    V_X=\frac{\delta v_x}{\epsilon \mathcal{V}_c}, \quad V_Y=\frac{v_y}{\epsilon \mathcal{V}_c}, \quad V_Z =\frac{v_z}{\mathcal{V}_c}, \quad P=\frac{p}{\mathcal{P}_c}.
\end{multline}
The characteristic velocity and pressure scales $\mathcal{V}_c$ and $\mathcal{P}_c$, respectively, are discussed below. 
Under this nondimensionalization, the leading-order terms (in $\epsilon$) left in iNS are \cite{WC19, WC20}}:
\begin{align}
    \frac{\partial V_X}{\partial X}+\frac{\partial V_Y}{\partial Y}+\frac{\partial V_Z}{\partial Z}&=0,\label{COM-O1}\displaybreak[3]\\
    -\frac{\partial P}{\partial X}&=0,\label{COLM-X-O1}\\
    -\frac{\partial P}{\partial Y}&=0,\label{COLM-Y-O1}\\
    \hat{Re}\left(V_X\frac{\partial V_Z}{\partial X}+V_Y\frac{\partial V_Z}{\partial Y}+V_Z\frac{\partial V_Z}{\partial Z}\right)&=-\frac{\partial P}{\partial Z}+\frac{\partial^2 V_Z}{\partial Y^2}\label{COLM-Z-O1}.
\end{align}

The fluid domain is defined as the deformed conduit: $\{(X,Y,Z)|-1/2<X<+1/2, 0<Y<H(X,Z), 0<Z<1\}$, in terms of the dimensionless variables. Here, $\hat{Re}$ is the modified Reynolds number, defined as $\hat{Re}=\epsilon Re =\epsilon \rho\mathcal{V}_c h_0/\mu$, where $\rho$ and $\mu$ are the fluid's density and dynamic viscosity, respectively. Equation~\eqref{COLM-Z-O1} relates the characteristic pressure and velocity scales as $\mathcal{P}_c = \mu \ell \mathcal{V}_c/h_0^2$. Equations \eqref{COLM-X-O1} and \eqref{COLM-Y-O1} indicate that, at the leading order in $\epsilon$ and $\delta$, the hydrodynamic pressure $P$ is only a function of the streamwise location $Z$, as in classical hydraulics problems \cite{W06_book}. Importantly, however, in this soft hydraulics problem, the hydraulic resistance (set by the cross-sectional shape and area \cite{W06_book,B08} via equation~\eqref{COLM-Z-O1}) is \emph{not} constant, and also varies with $Z$.

The following discussion begins with the case of $\hat{Re}\to 0$ (in section~\ref{sec:zero-Re}), \textit{i.e.}, flow with negligible inertia. In this case, we consider two different mechanical responses of the compliant microchannel's wall, for which analytical solutions, based on the notion of a slowly-varying \cite{VD87} unidirectional flow solution \footnote{Therefore, it is important to note that, unlike the case of unidirectional flows in rigid ducts, the solutions discussed herein are \emph{not} exact solutions of the incompressible Navier--Stokes equations \cite{LSS04}.}, are available in the literature \citep{CCSS17,SC18,WC19,AMC20,STGB17,BGB18}. For both types of mechanical response, the previous solutions yield a 3D model, in which the axial flow profile $V_Z=V_Z(X,Y,Z)$ and the top wall shape $H=H(X,Z)$ are coupled via the hydrodynamic pressure $P(Z)$. Our goal here is to first construct and validate reduced 2D models by ``removing'' the $X$ dependence in a suitably rigorous way, so that $H=H(Z)$ only. Upon accomplishing this reduction, averaging the 2D model over $Y$ yields a 1D model in which $H=H(Z)$ and $P=P(Z)$ are the remaining dependent variables. Therefore, when we extend the model to account for $\hat{Re}=\mathcal{O}(1)$ (in section~\ref{sec:nonzero-Re}), \textit{i.e.}, to flow with moderate inertia, it suffices to consider just one reduced model (instead of each mechanical response individually).


\section{Negligible flow inertia: \texorpdfstring{$\hat{Re}\to 0$}{hatRe->0}}
\label{sec:zero-Re}

\subsection{Effective deformed channel height}
\label{sec:heff}

Neglecting the inertia of the flow by taking $\hat{Re}\to0$ in equation~\eqref{COLM-Z-O1}, we find that the axial velocity $V_Z$, subject to the no-slip boundary condition at the walls, has a parabolic variation along the height of the duct ($Y$-direction):
\begin{equation}\label{VZ}
    V_Z(X,Y,Z)=-\frac{1}{2}\frac{\rd P}{\rd Z}Y[H(X,Z)-Y].
\end{equation}
At steady state, the flow rate is
\begin{equation}\label{Qint}
    Q := \int_{-1/2}^{+1/2}\int_{0}^{H(X,Z)}V_Z(X,Y,Z)\,\rd Y\,\rd X = const.,
\end{equation}
and thus the pressure gradient is found from equations~\eqref{VZ} and \eqref{Qint} to be
\begin{equation}\label{dP-Q-H}
    -\frac{\rd P}{\rd Z} = \frac{12Q}{\int_{-1/2}^{+1/2} H^3(X,Z)\,\rd X}.
\end{equation}

Equation~\eqref{dP-Q-H} can satisfy either one or two pressure boundary conditions (BCs). On the one hand, if the flow rate is controlled, then we can enforce $Q=q/q=1$ (\textit{i.e.}, take $\mathcal{P}_c=\mu q \ell/(wh_0^3)$ in the nondimensionalization) and set the outlet pressure to gauge, \textit{i.e.}, $P(Z=1)=0$. On the other hand, if the pressure drop $\Delta P = P(Z=0)-P(Z=1)$ is controlled, then enforcing $P(0)=p(0)/\Delta p=1$ (\textit{i.e.}, taking $\mathcal{P}_c=\Delta p$ in the nondimensionalization) is now also a BC, in addition to $P(Z=1)=0$, from which $Q$ is determined like an eigenvalue. Thus, in principle, the dimensionless flow rate $Q$ and the dimensionless pressure drop $\Delta P$ are not independent \cite{CCSS17}, and we do not specify the flow regime \textit{a priori} to make our results general. Either way, the pressure distribution in the duct can be determined {by integrating equation~\eqref{dP-Q-H} in $Z$, as long as the shape of fluid--solid interface, \textit{i.e.}, $H(X,Z)$, is known.}

Before we introduce expressions for $H(X,Z)$, recall that, in a wide rigid rectangular duct, the relation between the pressure gradient and the flow rate is set by a Poiseuille-like law \cite{W06_book}: 
\begin{equation}\label{dp-q-h-0}
    -\frac{\rd p}{\rd z} = \frac{12\mu q}{wh_0^3}.
\end{equation}
Thus, for a clearer comparison, it is helpful to transform equation \eqref{dP-Q-H} back into the dimensional form as
\begin{equation}\label{dp-q-h}
    -\frac{\rd p}{\rd z} = \frac{12\mu q}{\int_{-w/2}^{+w/2} h^3(x,z)\,\rd x}.
\end{equation}
In order to consistently rewrite equation \eqref{dp-q-h} in the form of a Poiseuille-like law~\eqref{dp-q-h-0}, we define the effective channel height as
\begin{equation}\label{heff}
    h_e(z) := \left[\frac{1}{w}\int_{-w/2}^{+w/2}h^3(x,z)\,\rd x\right]^{1/3}.
\end{equation}
Then, equation \eqref{dp-q-h} can be rewritten as
\begin{equation}\label{dp-q-heff}
    -\frac{\rd p}{\rd z} = \frac{12\mu q}{w h_e^3(z)}.
\end{equation}
Note that the corresponding dimensionless effective channel height is
\begin{equation}\label{Heff}
    H_e(Z) :=\frac{h_e(z)}{h_0}= \left[\int_{-1/2}^{+1/2}H^3(X,Z)\,\rd X\right]^{1/3}.
\end{equation}

Equation \eqref{dp-q-heff} can be viewed as a generalization of the Poiseuille-like law (for a wide rigid rectangular duct) {to a variable-height microchannel}. From another perspective, using the axially varying height $h_e(z)$ in equation~\eqref{dp-q-heff} eliminates the spanwise $x$-dependence of $h(x,z)$. Then, since the velocity was already averaged across the cross-section (to introduce $q$), the original 3D model has been reduced to an effective 1D model. Note that $h_e$ is meaningful only when speaking of the relation between $q$ and $\rd p/\rd z$, both of which only vary with $z$. This fact does \emph{not} mean that the velocity field is also 1D (it still depends on \emph{both} $y$ and $z$, thus remaining 2D). The effective height concept will be used to evaluate the accuracy of previous empirically-motivated reduced-order models.

In particular, in the original studies using 1D models, such as those proposed by \citet{GEGJ06} and \citet{HUZK09}, the \emph{average} deformed channel height
\begin{equation}\label{havg}
    \bar{h}(z) :=\frac{1}{w}\int_{-w/2}^{+w/2}h(x,z)\,\rd x
\end{equation}
is used in equation~\eqref{dp-q-heff} instead of $h_e(z)$. 
The corresponding dimensionless averaged channel height is
\begin{equation}\label{Havg}
    \bar{H}(Z) := \frac{\bar{h}(z)}{h_0} =\int_{-1/2}^{+1/2} H(X,Z)\, \rd X.
\end{equation}
It should be clear, however, that $\bar{h}$ (or $\bar{H}$) from equation~\eqref{havg} (or equation~\eqref{Havg}) is \emph{not} equal to $h_e$ (or $H_e$) from equation~\eqref{heff} (or equation~\eqref{Heff}). Importantly, the averaging approach (introducing $\bar{h}$ instead of $h_e$) leads to an inconsistency in the reduced model because if we replace $h_e^3(z)$ with $\bar{h}^3(z)$ in equation~\eqref{dp-q-heff}, then it is no longer equivalent to equation~\eqref{dp-q-h}, which was rigorously derived by integrating the leading-order iNS \eqref{COM-O1}--\eqref{COLM-Z-O1}. In the present work, our goal is to determine how this inconsistency affects the hydraulic predictions.

To finish the derivation, we must specify $H(X,Z)$. {In the present context of soft hydraulics, $H(X,Z)$ is determined by solving an appropriate solid mechanics (elasticity) problem.} Our assumption that the { deformed} microchannel { remains} long and shallow { ($h(x,z)\ll w\ll\ell$) so that $u_y\ll w$ (recall that $h_0\ll w$ and $h(x,z) = h_0 +u_y(x,z)$). If the top wall is thick enough, with $w \lesssim t \ll \ell$, then the shallowness and slenderness of the deformed channel enforces} small-strain deformation and allows the use of linear elasticity. { If the top wall is thin with $t\lesssim w\ll \ell$, we require that $\max_{x,z} u_{y}\ll t$ to make the linear elastic theory applicable \cite{CCSS17, SC18, AMC20}. However, regardless of the wall thickness, as long as $t\ll \ell$, the original 3D elasticity problem can be reduced to a 2D one. Here, we only briefly outline the reasons for the statement, and the reader is directed to Ref.~\onlinecite{WC19} for the detailed analysis. First, the lubrication approximation implies \cite{W06_book} that the shear stress $\tau_{yz}\ll p$. Since the tractions are continuous across the fluid--solid interface, we can thus infer that $\sigma_{yz}\ll \sigma_{yy}$ where $\sigma_{yz}$ and $\sigma_{yy}$ are components of the Cauchy stress. Then, by examining the momentum balance in the solid, it is concluded that the dominant components of stress are in the cross-sectional $(x,y)$ plane. As a consequence, the deformation profiles at different streamwise ($z$-locations) decouple from each other, leading to a local deformation--pressure relation.} 

Now, from equations~\eqref{COLM-X-O1} and \eqref{COLM-Y-O1}, $P=P(Z)$ only, thus $P$ acts uniformly at each axial $(X,Y)$ cross-section to deform the top wall. Therefore, the spanwise deformation is determined by the local pressure $P(Z)$, and we can express the deformed duct shape as
\begin{equation}\label{H(X,Z)}
    H(X,Z) = \frac{h_0 + u_y(x,z)}{h_0}= 1 + \lambda F(X) P(Z),
\end{equation}
where $\lambda := \mathcal{U}_c/h_0$, with $\mathcal{U}_c$ being the characteristic displacement of the top wall, is a dimensionless group that captures the compliance the top wall. Restating the above-mentioned slenderness assumptions, we must require that $\lambda\ll 1/\delta$ for lubrication theory and linear elasticity to be applicable. The spanwise profile $F(X)$ is obtained by solving the corresponding elasticity problem in the $(X,Y)$ cross-section of the duct \cite{CCSS17,SC18,WC19,AMC20}. {Also, note that equation \eqref{H(X,Z)} is \emph{not} an assumption but a consequence of the asymptotic reduction of the elasticity problem for a long and slender microchannel. Since the analysis (summarized above) only involves balancing the momentum equation in the solid, it holds for any boundary conditions. However, the boundary conditions play a role in determining the actual deformation field, leading to different expressions for $F(X)$}.

Note that equation \eqref{H(X,Z)} takes the form of the deformation of a soft interface on a Winkler foundation \cite{W67,DMKBF18}, but now the foundation's (dimensionless) ``spring stiffness'' is given by $\lambda F(X)$. Winkler-foundation-like relations between pressure and deformation arise in a number of soft lubrication problems \cite{SM04,SM05}, including particles near elastic substrates \cite{CC10,KCC20}, slider bearings \cite{CC11}, and rollers \cite{YK05}. The analogy becomes even stronger upon introducing the concept of averaged deformed channel height. Specifically, substituting equation \eqref{H(X,Z)} into equations \eqref{Heff} and \eqref{Havg}, respectively, we obtain explicit expressions for $H_e(Z)$ and $\bar{H}(Z)$ as
\begin{equation}\label{Heff2}
\begin{aligned}
    H_e(Z) = \big[1 &+ 3\mathcal{I}_1\lambda P(Z) \\
    &+ 3\mathcal{I}_2\lambda^2 P^2(Z) + \mathcal{I}_3\lambda^3 P^3(Z) \big]^{1/3},
\end{aligned}
\end{equation}
and
\begin{equation}\label{Havg2}
    \bar{H}(Z) = 1 + \mathcal{I}_1\lambda P(Z).
\end{equation}
Then, from equation~\eqref{Havg2} the now-constant (dimensionless) spring stiffness in the analogy to a Winkler foundation is $\xi = \mathcal{I}_1\lambda$. Here, the coefficients $\mathcal{I}_i$ are defined as
\begin{equation}\label{Ii}
    \mathcal{I}_i := \int_{-1/2}^{+1/2}F^i(X)\,\rd X, \quad i=1,2,\hdots.
\end{equation}

Interestingly, observe that $\bar{H}$ in equation~\eqref{Havg2} is simply the one-term Taylor-series approximation of $H_e$ from equation~\eqref{Heff2} in terms of $\lambda \ll 1$. However, our analysis does not require $\lambda \ll 1$, in fact $\lambda=\mathcal{O}(1)$ is possible. Linear elasticity only requires that  $\lambda\ll 1/\delta$ (as discussed by \citet{WC19} and \citet{SC18}). Thus, we would like to determine if the approximation in going from equation \eqref{Heff2} to equation \eqref{Havg2} is a valid one. 

\subsection{Flow rate--pressure drop relation}

To obtain the general form of the flow rate--pressure drop relation in a soft hydraulic conduit, we return to the dimensionless form of equation \eqref{dP-Q-H}, namely:
\begin{equation}\label{dP-Q-Heff}
    -\frac{\rd P}{\rd Z} = \frac{12Q}{ H_e^3(Z)}.
\end{equation}
Since $Q=const.$ in steady flow, upon substituting equation \eqref{Heff2} into equation \eqref{dP-Q-Heff}, we obtain a separable first-order ordinary differential equation (ODE) for $P(Z)$. The solution, subject to $P(1)=0$, is 
\begin{align}\label{QP-Heff}
    12Q(1-Z) &= P(Z)\left[1+\frac{3}{2}\mathcal{I}_1\lambda P(Z)+\mathcal{I}_2\lambda^2P^2(Z) \right. \nonumber\\
    &\qquad\qquad\quad \left.+\frac{1}{4}\mathcal{I}_3\lambda^3P^3(Z)\right].
\end{align}

As discussed in section~\ref{sec:heff}, previous empirical studies used $\bar{H}$ in place of $H_e$. In this case, substituting equation \eqref{Havg2} into equation \eqref{dP-Q-Heff}, and solving the corresponding ODE, yields an explicit expression for the pressure distribution: 
\begin{align}\label{QP-Havg}
    P(Z) = \frac{1}{\mathcal{I}_1\lambda}\left\{\left[48\mathcal{I}_1\lambda Q(1-Z)+1\right]^{1/4}-1\right\}.
\end{align}
As mentioned in section~\ref{sec:heff}, we may either consider a flow-controlled situation, in which $Q=1$ and $\Delta P=P(0)$ is found implicitly from equation~\eqref{QP-Heff} or explicitly from equation \eqref{QP-Havg}. Meanwhile in the pressure-controlled regime, we enforce $P(0)=1$ and compute $Q$ directly:
\begin{equation}\label{eq:Q_Re0}
    Q = \frac{1}{48}\times\begin{cases}
    4 + 6\mathcal{I}_1\lambda + 4\mathcal{I}_2\lambda^2 + \mathcal{I}_3\lambda^3,\quad &\text{from }\eqref{QP-Heff},\\[2mm]
    \displaystyle\frac{1}{\mathcal{I}_1\lambda}\left[(\mathcal{I}_1\lambda + 1)^4-1\right],\quad &\text{from }\eqref{QP-Havg}.
    \end{cases}
\end{equation}

Equation \eqref{QP-Havg} is essentially the same model derived by \citet{GEGJ06}. However, in said work,  $\mathcal{I}_1\lambda$ was taken as an unknown parameter, denoted as $\alpha$, which was calibrated against experiments. However, our equation \eqref{QP-Havg} is parameter-free because both $\lambda$ and $\mathcal{I}_1$ are known from solving a suitable elasticity problem. Therefore, our approach eliminates the ambiguity, pointed out by \citet{HUZK09}, of what unknown dependencies ``hide'' in $\alpha$.

Note, however, that even if equation \eqref{Havg2} is the one-term Taylor-series approximation to \eqref{Heff2}, this is not true for the flow rate--pressure drop relations  \eqref{QP-Havg} and \eqref{QP-Heff}, respectively. Therefore, we must determine how well $P(Z)$ based on the averaged channel height approximates $P(Z)$ based on the effective channel height. It is reasonable to conjecture that, due to the restriction to small strains required by linear elasticity, the two expressions should be in close agreement. To substantiate this conjecture, we proceed to quantify the difference between equations \eqref{QP-Heff} and \eqref{QP-Havg} to obtain insight into the error committed in the formulation based on the averaged channel height. To this end, we apply the methodology established in this subsection to two types of common microchannel wall deformations considered in the literature: a microchannel with a thick top wall (section~\ref{sec:Re0_thick}) and a microchannel with a thinner, plate-like top wall (section~\ref{sec:Re0_plate}).

\subsection{Illustrated examples}

\subsubsection{Duct with thick compliant top wall}
\label{sec:Re0_thick}

First, we analyze the case of an initially rectangular duct with three compliant walls embedded in a thick soft structure. The channel's shallowness makes the deformation of the side wall negligible compared with that of the top wall. Thus, the schematic diagram in figure~\ref{fig:schematic} still applies. The corresponding steady 3D FSI problem was solved by \citet{WC19}. To summarize their key conclusions: although a solution was obtained for any $t/w$, it was shown that, for $t/w\gtrsim 1.5$, the ``thick'' limit ($t^2/w^2\gg1$) is achieved and a simple analytical Fourier series solution can be written down for the deformed channel's top wall:
\begin{align}
    h(x,z)&=h_0\left[1+\frac{wp(z)}{\Bar{E}h_0}\mathfrak{f}(x)\right] \label{eq:h_f_thick}\\
    \mathfrak{f}(x)&=\sum_{m=1}^{\infty}\frac{2A_m}{m\pi}\sin\left[m\pi\left(\frac{x}{w}+\frac{1}{2}\right)\right],
    \label{eq:frak_f_thick}
\end{align}
where we have defined $A_m:=\frac{2}{m\pi}[1-(-1)^m]$ and  $\Bar{E}:=E/(1-\nu^2)$, with $E$ being Young's modulus and $\nu$ the Poisson's ratio.

From equation~\eqref{eq:frak_f_thick}, we can determine the function $F(X)\equiv F(x/w) = \mathfrak{f}(x)$ introduced in equation~\eqref{H(X,Z)}. The corresponding values of $\mathcal{I}_i$, defined in equation~\eqref{Ii}, are computed and summarized in table \ref{tab:table-I}.

\begin{table}[hb]
    \caption{\label{tab:table-I} Values of the coefficients $\{\mathcal{I}_i\}_{i=1}^3$ defined by equation~\eqref{Ii} for the thick-walled microchannel.}
    \begin{tabular}{l @{\hspace{3em}} c}
    \toprule
    $\mathcal{I}_1$ & 0.542754 \\ 
    \hline
    $\mathcal{I}_2$ & 0.333333 \\
    \hline
    $\mathcal{I}_3$ & 0.215834 \\
    \botrule
    \end{tabular}
\end{table}

\begin{figure*}
    \centering
    \includegraphics[width = 0.75\textwidth
    ]{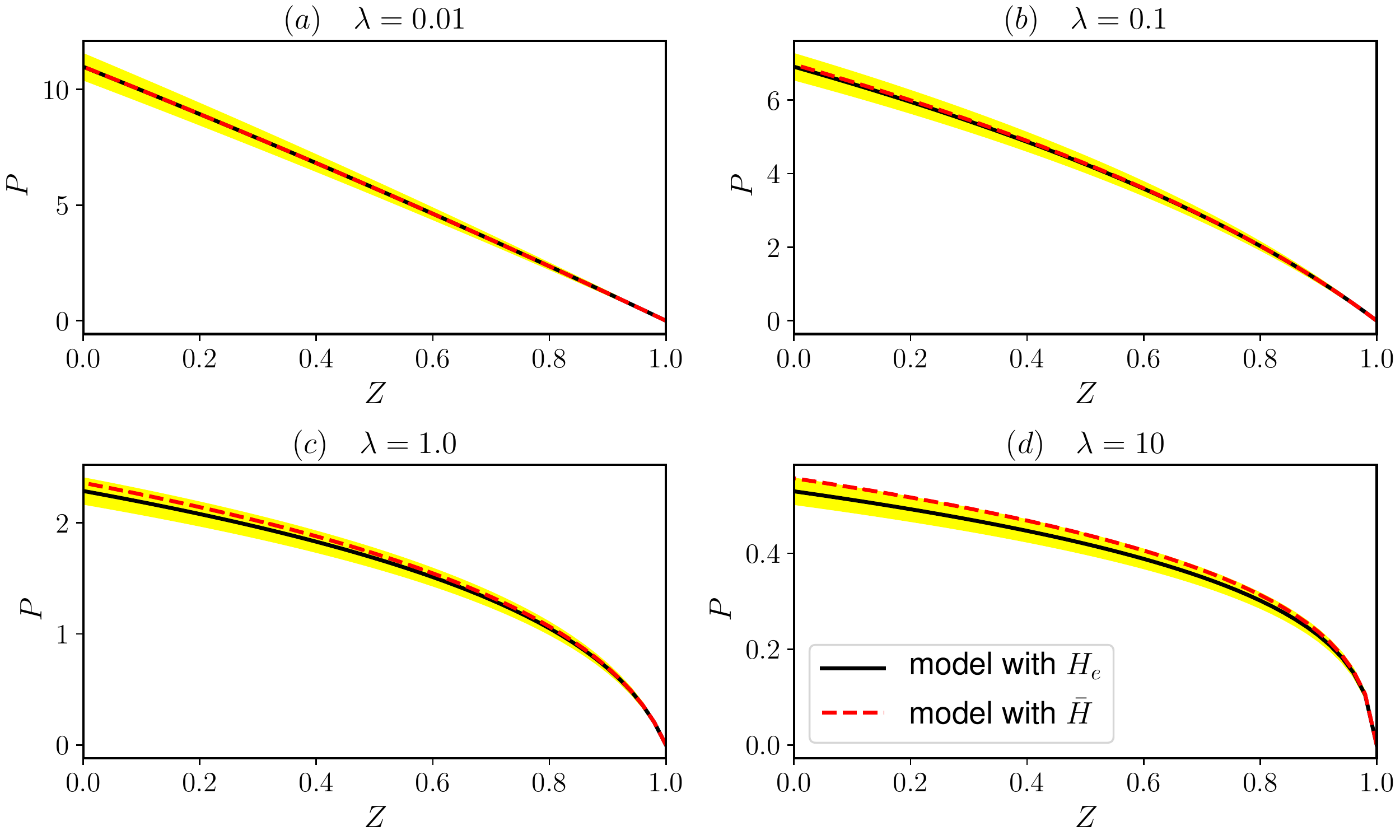}
    \caption{Thick top wall: Axial pressure distribution $P(Z)$ in a soft hydraulic conduit for $Q=1$ and different $\lambda$: (a) $\lambda = 0.01$, (b) $\lambda=0.1$, (c) $\lambda=1.0$, and (d) $\lambda=10$. The solid curve is computed from equation \eqref{QP-Heff}, in which the effective channel height \eqref{Heff} is employed, while the dashed curve is computed from equation \eqref{QP-Havg}, in which the averaged channel height \eqref{Havg} is employed. The shaded region represents $\pm 5\%$ of deviation from the solid curve, which is the baseline (or ``truth'') for this model.}
    \label{fig:PZ-lam-thick}
\end{figure*}

The compliance parameter $\lambda$ emerges naturally from the nondimensionalization of equation \eqref{eq:h_f_thick}:
\begin{equation}\label{lam-thick}
    \lambda = \frac{w\mathcal{P}_c}{h_0\bar{E}} = \begin{cases} \displaystyle\frac{\mu q\ell}{h_0^4\bar{E}} &~~\text{(flow controlled)},\\[5mm]
    \displaystyle\frac{w\Delta p}{h_0\bar{E}} &~~\text{(pressure controlled)}.\end{cases}
\end{equation}

Substituting $\lambda$ and $\mathcal{I}_i$ into equation \eqref{QP-Heff} and \eqref{QP-Havg} respectively, we are ready to make a comparison between the two formulations. We observe that the pressure distribution depends nonlinearly upon $\lambda$, as illustrated in figure~\ref{fig:PZ-lam-thick}. The total pressure drop $\Delta P=P(0)$ decreases with $\lambda$, and a strong  pressure gradient develops near the outlet. Notably, even with $\lambda$ varying by three orders, the results computed with the two equation remain close to each other. The pressure distribution computed from equation \eqref{QP-Havg}, which employs the averaged channel height, is slightly higher than that from equation \eqref{QP-Heff}, which employs the effective channel height. However, the difference is no larger than $5\%$ for almost the whole range of $\lambda$ values considered. (The maximum deviation is found to be $5.14\%$ in the case of $\lambda=10$, which is pushing the limit of the applicability of the theory.)  Having computed $P(Z)$, $H_e(Z)$ and $\bar{H}(Z)$ can be found from equations~\eqref{Heff} and  \eqref{Havg}, respectively. The largest deformed height is at the channel inlet (\textit{i.e.}, at $Z=0$), and we can expect the approximation of the effective channel height by the averaged one to be worst there. However, we determined that $\max_{0\le\lambda\le10} |H_e(0)-\bar{H}(0)|/H_e(0) < 5\%$, showing good agreement.

Now that the validity of the approximate prediction of the flow rate--pressure drop relation \eqref{QP-Havg}  has been established, it is worthwhile to provide a formula for the fitting parameter $\alpha$ introduced by \citet{GEGJ06}. Recall the averaged channel height from the latter model is
\begin{equation}\label{havg-Gervais}
    \bar{h}(z) = h_0\left[1+\alpha\frac{wp(z)}{Eh_0}\right].
\end{equation}
For a clearer comparison, we transform equation \eqref{Havg} into its dimensional form:
\begin{equation}\label{havgthick}
    \Bar{h}(z) = h_0\left[1+\mathcal{I}_1(1-\nu^2)\frac{wp(z)}{Eh_0}\right].
\end{equation}
Then, comparing equations \eqref{havg-Gervais} and \eqref{havgthick}, it is readily recognized that 
\begin{equation}\label{eq:alpha-thick}
    \alpha = \mathcal{I}_1(1-\nu^2)\approx 0.542754(1-\nu^2),
\end{equation}
which we observe is a function of the Poisson's ratio, but no other material or geometric parameters related to the top wall, in this thick-wall limit ($t^2/w^2\gg1$). (This observation will be contrasted with the result in equation~\eqref{eq:alpha-plate} below.) Furthermore, most microchannels are made from materials such as polydimethylsiloxane (PDMS) \cite{MW02,FY10}, which is often considered a nearly incompressible material, \textit{i.e.}, $\nu\approx0.5$. Then, $\alpha\approx 0.4071$. A different solid mechanics model (and response) for the top wall would yield a different estimate of $\alpha$ (see section~\ref{sec:Re0_plate}), showing that $\alpha$ is \emph{not} a universal number that can be determined by a single set of experiments (even if this approach works for some sets of geometries). Nevertheless, equation~\eqref{eq:alpha-thick} provides a quantitative connection between the earlier scaling models \cite{GEGJ06} for flow-induced deformation and the later detailed elasticity calculations \cite{WC19}.

It is also relevant to mention that the results in this subsection also yield insight into the quality of approximation of another approach to the flow-induced deformation problem. For example, following \citet{SM04,SM05} and \citet{CC10}, \citet{MCC13} expressed the deformation at the fluid--solid interface of a thick-walled 2D duct as 
\begin{equation}\label{eq:h_2D}
    h(z) = h_0\left[1 + \frac{H_1p(z)}{h_0 E_m}\right],
\end{equation}
where the layer thickness $H_1$ and its ``effective'' Young's modulus $E_m$ can be  considered adjustable parameters \footnote{Such models have been found useful in analyzing the global inflation or relaxation time scale of a microchannel, which is relevant to the start-up problem and stop-flow lithography \cite{DGPHD07,PYDHD09}.}. In particular, $H_1$ represents the distance over which the vertical displacement varies, vanishing at $y=H_1$. Equation~\eqref{eq:h_2D} is based on assuming no spanwise variation, reducing the flow and deformation problem to a 2D setting in the $(y,z)$ plane, thus $h=h(z)$ \emph{a forteriori} now (no averaging). The obvious question that arises is: what are suitable values of $H_1$ and $E_m$? As with equation~\eqref{havg-Gervais}, we simply compare equation~\eqref{eq:h_2D} to \eqref{havgthick} to obtain the answer. We conclude that 
\begin{equation}\label{eq:H1-Em-thick}
   \frac{H_1}{E_m} = \mathcal{I}_1(1-\nu^2)\frac{w}{E} \approx 0.542754(1-\nu^2)\frac{w}{E}.
\end{equation}
For example, if the 2D soft layer is taken to have the same elastic properties as the 3D one it approximates, $E_m=E$, then equation~\eqref{eq:H1-Em-thick} provides its suitable thickness $H_1$ as a function of $\nu$ and $w$. Note that, separately, \citet{EPKVS21} surveyed a number of such two-dimensional elastohydrodynamic problems, while \citet{CV20} critically addressed the 2D models' validity in the near-incompressible limit as $\nu\to1/2^{-}$.

\begin{figure*}[ht]
    \centering
    \includegraphics[width=0.75\textwidth]{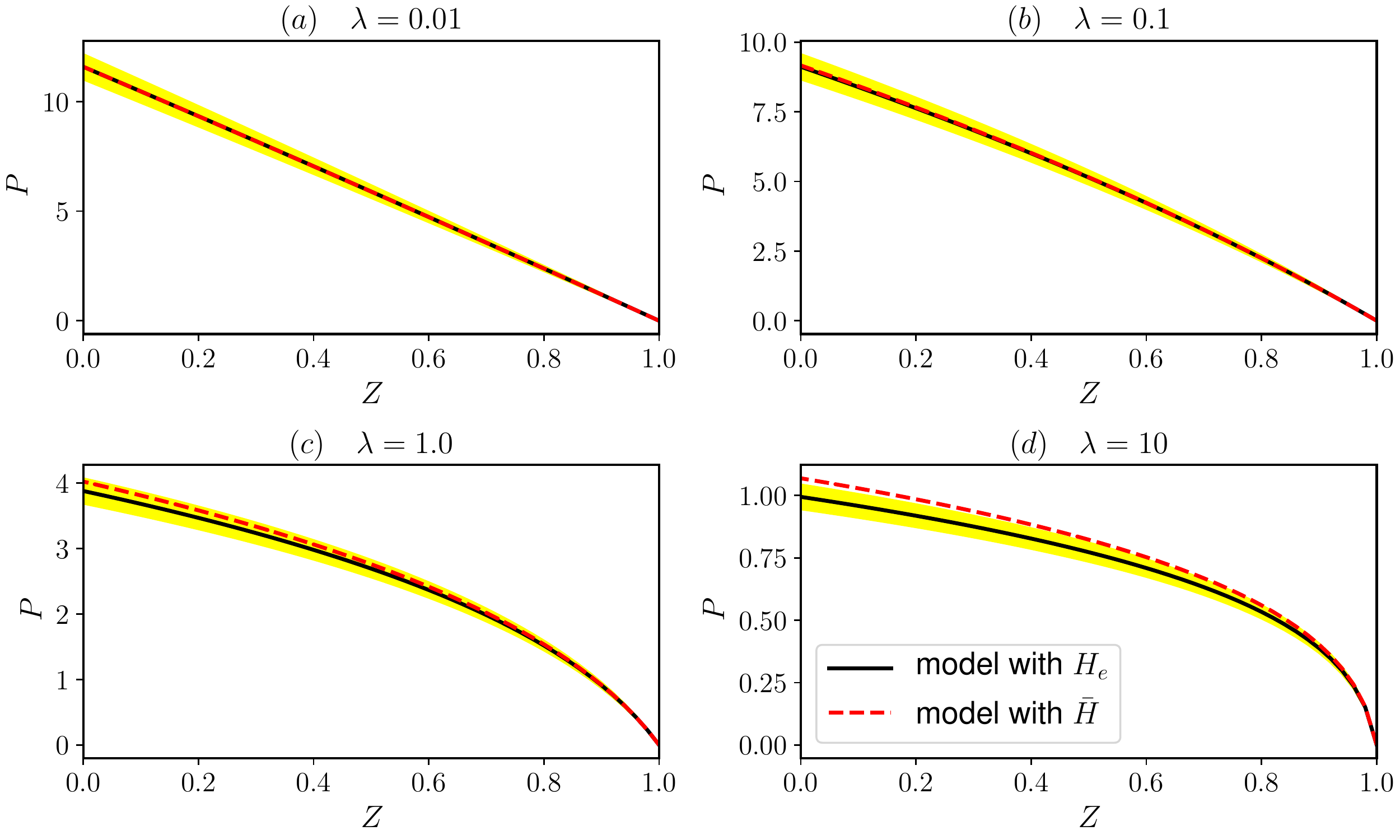}
    \caption{Plate-like top wall: Axial pressure distribution $P(Z)$ in a soft hydraulic conduit for $Q=1$ and different $\lambda$: (a) $\lambda = 0.01$, (b) $\lambda=0.1$, (c) $\lambda=1.0$, and (d) $\lambda=10$. The solid curve is computed from equation \eqref{QP-Heff}, in which the effective channel height \eqref{Heff} is employed, while the dashed curve is computed from equation \eqref{QP-Havg}, in which the averaged channel height \eqref{Havg} is employed. The shaded region represents $\pm 5\%$ of deviation from the solid curve, which is the baseline (or ``truth'') for this model. The top wall thickness-to-width ratio is $t/w=0.5$.}
    \label{fig:PZ-lam-plate}
\end{figure*}

\subsubsection{Duct with plate-like compliant top wall}
\label{sec:Re0_plate}

Next, we analyze the case of a duct with a clamped thick-plate-like compliant top wall. As in section~\ref{sec:Re0_thick}, the slenderness of the duct still results in the decoupling of the top wall deformation at each streamwise cross-section. However, the resulting shape of the deformed fluid--solid interface obtained by \citet{SC18} is quite different from equations~\eqref{eq:h_f_thick}--\eqref{eq:frak_f_thick}. Specifically, now
\begin{align}
    h(x,z) &= h_0\left[ 1 + \frac{w^4p(z)}{24Bh_0} \mathfrak{f}(x) \right] \label{eq:h_f_plate},\\
    \mathfrak{f}(x) &=  \left[\frac{1}{4}-\left(\frac{x}{w}\right)^2\right] \left\{\frac{2(t/w)^2}{\kappa(1-\nu)}+ \left[\frac{1}{4}-\left(\frac{x}{w}\right)^2\right]\right\},
    \label{eq:frak_f_plate}
\end{align}
where $B=\Bar{E} t^3/12$ is the plate's flexural rigidity \cite{TWK59}, and $\kappa$ is the ``shear correction factor'' \cite{CE19}. For consistency with the theory of elasticity, $\kappa=1$ should be imposed \cite{Z06}, but we leave it in the equations for the sake of completeness. The plate model considers bending deformation, as well as shear deformation, of the top wall, and it is applicable for $t\lesssim w$. If $t^2/w^2\ll 1$, the first term in the inner curly brace in equation~\eqref{eq:frak_f_plate} is negligible, meaning the shear deformation is not important in this case. The model then reduces to the one derived earlier by \citet{CCSS17}, which only accounted for plate bending.

By making equation \eqref{eq:h_f_plate} dimensionless, we obtain
\begin{equation}\label{lam-plate}
    \lambda = \frac{w^4\mathcal{P}_c}{24h_0B} = 
    \begin{cases}\displaystyle\frac{\mu qw^3\ell}{24h_0^4B} &~~\text{(flow controlled)},\\[5mm]
    \displaystyle\frac{w^4\Delta p}{24h_0B} &~~\text{(pressure controlled)}.\end{cases}
\end{equation}
Again, we have $F(X) \equiv F(x/w) = \mathfrak{f}(x)$, but $\mathfrak{f}$ is now given by equation~\eqref{eq:frak_f_plate}.
Then, equation \eqref{eq:h_f_plate} takes the same form as equation \eqref{H(X,Z)}. Next, the calculation of the $\mathcal{I}_i$ can be done explicitly for this case, yielding the functions of $t/w$, $\kappa$ and $\nu$ summarized in table~\ref{tab:table-II}.

\begin{table}
    \caption{\label{tab:table-II} Functional forms of the coefficients $\{\mathcal{I}_i\}_{i=1}^3$ defined by equation~\eqref{Ii} for the plate-like-walled microchannel.}
    \begin{tabular}{l @{\qquad} l}
    \toprule
    $\mathcal{I}_1$ & $\frac{1}{30}+\frac{(t/w)^2}{3\kappa(1-\nu)}$\\
    \hline
    $\mathcal{I}_2$ & $\frac{1}{630}+\frac{(t/w)^2}{35\kappa(1-\nu)} + \frac{2(t/w)^4}{15\left[\kappa(1-\nu)\right]^2}$\\
    \hline
    $\mathcal{I}_3$ & $\frac{1}{12012}+\frac{(t/w)^2}{462\kappa(1-\nu)} + \frac{2(t/w)^4}{105\left[\kappa(1-\nu)\right]^2} + \frac{2(t/w)^6}{35\left[\kappa(1-\nu)\right]^3}$\\
    \botrule
    \end{tabular}
\end{table}

As in section \ref{sec:Re0_thick}, we now substitute equation \eqref{lam-plate} into equations \eqref{Heff} and \eqref{Havg} respectively and compare the results. Figure~\ref{fig:PZ-lam-plate} shows $P(Z)$ for different $\lambda$ and $Q=1$. The two formulations predict similar results. The error committed by replacing $H_e$ with $\Bar{H}$ is $< 8\%$. However, even with  smaller or larger $t/w$ ratios, the pressure distributions computed with each $H$ expression do not differ much from each other. The maximum deviation is $<9\%$. As in section~\ref{sec:Re0_thick}, we computed the absolute difference between using $H_e(0)$ and $\bar{H}(0)$, and found that $\max_{0\le\lambda\le10} |H_e(0)-\bar{H}(0)|/H_e(0) < 5\%$.

Finally, we can also compare the model \eqref{QP-Havg} (formulated with the averaged channel height) to equation \eqref{havg-Gervais} (the model derived by \citet{GEGJ06}) to obtain an explicit expression for the fitting parameter $\alpha$. Again, for convenience, we write the dimensional form of the averaged channel height as
\begin{equation}\label{havgplate}
\begin{aligned}
    \Bar{h}(z) &= h_0\left[1+\mathcal{I}_1\frac{w^4p(z)}{24Bh_0}\right]\\
    &=h_0\left[1+\mathcal{I}_1\left(\frac{1-\nu^2}{2}\right)\left(\frac{w}{t}\right)^3\frac{wp(z)}{E h_0}\right].
\end{aligned}    
\end{equation}
It follows, in this case, that
\begin{equation}\label{eq:alpha-plate}
    \begin{aligned}
    \alpha &= \mathcal{I}_1\left(\frac{1-\nu^2}{2}\right)\left(\frac{w}{t}\right)^3 \\ 
    & = \left(\frac{1-\nu^2}{60}\right)\left[\left(\frac{w}{t}\right)^3 + \frac{10}{\kappa(1-\nu)}\left(\frac{w}{t}\right)\right].
\end{aligned}
\end{equation}
Observe that, unlike equation~\eqref{eq:alpha-thick}, $\alpha$ now depends upon $w$ and $t$ (with $w/t\gtrsim1$), in addition to $\nu$. The dependence on $t$, which equation \eqref{eq:alpha-plate} now quantitatively predicts, has been observed in experimental studies \cite{HUZK09,RDC17}.

\subsection{A fiction factor for laminar flow in compliant ducts}
\label{sec:friction_factor}

Recently, it has been of interest to extend the textbook notion of a friction factor for various flows in microchannels. One idea is to take into account the shear-rate-dependent viscosity of non-Newtonian fluids \cite{ME08}. Even for Newtonian fluids, updates are being sought to better understand (the previously considered ``settled'') wall roughness effects in both the laminar \cite{LLS19} and turbulent \cite{F18} portions of the Moody diagram (the visual representation of the friction factor \cite{M44}). A friction factor is needed for microfluidic system design \cite{SASM01}, much like its use for analyzing pipe networks \cite{P11}. A frontier application is \emph{microrheometry} \cite{PM09,GWV16}, in which an experimentally computed friction factor in a rectangular microchannel is compared to a theoretical value, in order to characterize the viscosity of a fluid \cite{YWHFJZM19}. An open problem in microrheometry \citep{DGNM16} concerns whether measurements made in PDMS microchannels are affected by the friction factor's implicit $\Delta p/E$ (or, in the present notation, $\lambda$) dependence. As the discussion above makes clear, the deformation of a compliant duct indeed changes the pressure drop characteristics. Thus, a salient application of our reduced-order flow and deformation model from section~\ref{sec:zero-Re} is to interrogate the dependence of the friction factor on the elasticity-related parameters and variables.

To this end, we start from the reduced model with the averaged channel height as the effective channel height, \textit{i.e.}, $h_e(z) =\bar{h}(z)=h_0[1+\eta p(z)]$. Note the compliance constant $\eta=\xi/\mathcal{P}_c$, with $\xi=\lambda\mathcal{I}_1$ being the dimensionless spring stiffness parameter introduced in section~\ref{sec:heff}, is known from having solved a suitable solid mechanics problem. Then, from equation \eqref{QP-Havg}, we have
\begin{equation}\label{eta-p}
    \eta p(z) = \xi P(Z) = [48\xi (1-z/\ell)+1]^{1/4}-1,
\end{equation}
where we have substituted $Q=1$ and $Z=z/\ell$. Equation \eqref{eta-p} indicates that $\eta p$ cannot be varied independently because it is fully determined by $\xi$. In the following discussion, we work with dimensional variables for convenience.

For $\hat{Re}\to0$, the pressure difference across an axial length of a duct is balanced by the viscous drag on the wall. Denote the area of the cross section as $a(z)=w\bar{h}(z)$, which takes into account the area change due to the deformation of the top wall. Then, the mean shear stress \cite{W06_book} can be written as
\begin{equation}\label{tauw}
\begin{split}
    \bar{\tau}_w &= -\frac{1}{c_p}\left(\frac{\rd p}{\rd z}a+p\frac{\rd a}{\rd z}\right)\\
    &= -\frac{wh_0}{c_p} \left(\frac{\rd p}{\rd z}(1+\eta p)+\eta p\frac{\rd p}{\rd z}\right)\\
    &= \frac{D_{h_0}}{4}(1+2\eta p)\left(-\frac{\rd p}{\rd z}\right)\\
    &= \frac{D_{h}}{4}\left(-\frac{\rd p}{\rd z}\right).
\end{split}
\end{equation}
Here, $c_p = 2(w+\bar{h})$ is the perimeter of the cross-section, and $c_p \approx 2(w+h_0)$ for $\bar{h}\ll w$. Additionally, $D_{h_0} = 4h_0w/[2(w+h_0)]$ is the \emph{hydraulic diameter} of a rigid rectangular duct \cite{W06_book}. In the last equality in equation~\eqref{tauw}, we further defined the hydraulic diameter of the soft duct as
\begin{equation}
    D_h := D_{h_0}(1+2\eta p),
\end{equation}
where $\eta p$ captures the flow-induced deformation, meaning that $D_h$ varies along the streamwise direction with $p$.

Next, consider the Fanning friction factor defined \cite{W06_book} as:
\begin{equation}\label{friction_factor1}
    \begin{split}
        C_f &:= \frac{2\bar{\tau}_w}{\rho\bar{v}_z^2}\\
        &= \frac{1}{2}D_h^2\left(-\frac{\rd p}{\rd z}\right) \left(\frac{\mu}{\rho\bar{v}_zD_h}\right) 
        \left(\frac{1}{\mu \bar{v}_z}\right)\\
        &= 6\left(\frac{D_h}{\bar{h}}\right)^2\frac{1}{Re_{D_h}},
    \end{split}
\end{equation}
where we have substituted equation \eqref{dp-q-heff} with $h_e=\bar{h}$ into the last step above. Also note that we have introduced the averaged velocity as $\bar{v}_z=q/(w\bar{h})$ and the hydraulic-diameter-based Reynolds number as 
\begin{equation}
    Re_{D_h}=\frac{\rho\bar{v}_zD_h}{\mu}=Re_{D_{h_0}}\left(1+\frac{\eta p}{1+\eta p}\right),
\end{equation}
with $Re_{D_{h_0}} = \rho qD_{h_0}/(\mu wh_0)$ being the Reynolds number for the rigid rectangular duct. 

Equation \eqref{friction_factor1} has a form similar to the friction factor for a rigid rectangular duct. However, all three parameters, $D_h$, $\bar{h}$ and $Re_{D_h}$, depend on $z$ due to FSI. To highlight this effect, we can re-write equation \eqref{friction_factor1} as
\begin{equation}\label{friction_factor2}
    C_f = \underbrace{6\left(\frac{D_{h_0}}{h_0}\right)^2\frac{1}{Re_{D_{h_0}}}}_{\text{rigid duct }C_f}\left(1+\frac{\eta p}{1+\eta p}\right).
\end{equation}
The first term in equation~\eqref{friction_factor2} is $C_f$ for a rigid rectangular duct, while the second term (in the parentheses) above captures the soft hydraulic effect. Furthermore, we can define the Poiseuille number as
\begin{equation}\label{eq:Po}
    Po := C_f Re_{D_h} = \underbrace{6\left(\frac{D_{h_0}}{h_0}\right)^2}_{\text{rigid duct }Po} \left(1+\frac{\eta p}{1+\eta p}\right)^2.
\end{equation}
We re-iterate that equation~\eqref{eq:Po} is valid only for $h_0\ll w$, and observe that the prefactor $6(D_{h_0}/h_0)^2=24$ for $h_0/w\to0$. Furthermore, while $Po=const.$ in a non-circular \emph{rigid} duct \cite{W06_book}, $Po$ from equation~\eqref{eq:Po} becomes a function of $z$ due to FSI. Further, the increase of the soft hydraulic $Po$ is clearly demonstrated by the second term in the last parenthesis on the right-hand side of equation~\eqref{eq:Po}, which is bounded between $1$ (as $\eta p\to0$) and $4$ (as $\eta p\to\infty$).

\begin{figure}[t]
    \centering
    \includegraphics[width=\columnwidth]{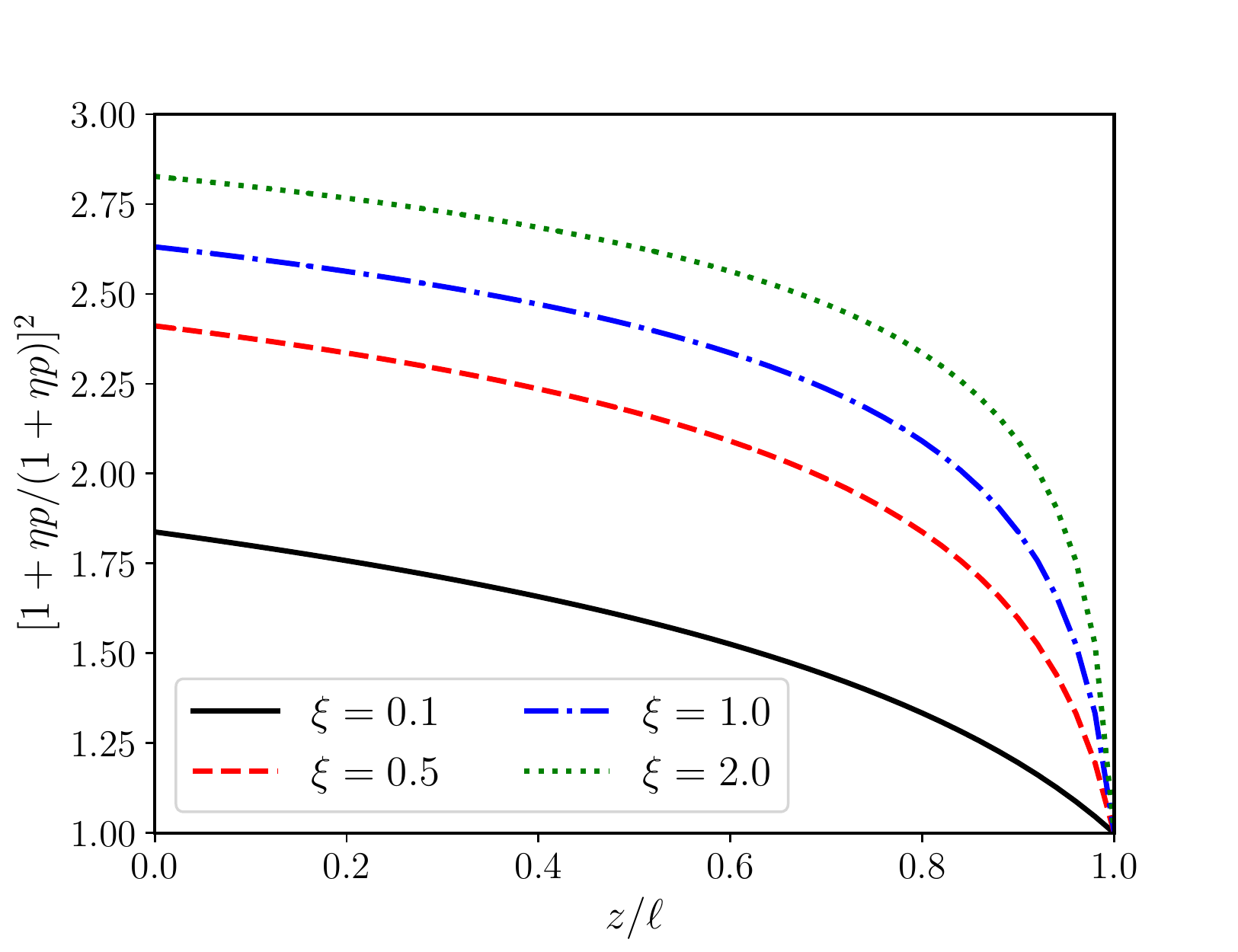}
    \caption{The variation of the reduced Poiseuille number $Po/(\text{rigid duct }Po)=[1+\eta p/(1+\eta p)]^2$ from equation~\eqref{eq:Po} along the flow wise direction, $z$, for different $\xi$.}
    \label{fig:po-xi}
\end{figure}

We highlight the novel dependence of $Po$ on the compliance parameter $\xi$, beyond the usual geometric dependence on $(D_{h_0}/h_0)^2$, by plotting $Po$ versus $z/\ell$ for given $\xi$, after eliminating $\eta p$ via equation \eqref{eta-p}. As predicted by equation \eqref{eq:Po}, figure~\ref{fig:po-xi} shows that $Po$ in a compliant duct is not a constant but rather a \emph{decreasing} function along the streamwise direction (since $p(z)$ is as well). The shape is strongly influenced by the value of $\xi$, even if ultimately the correction factor due to compliance is bounded between $1$ and $4$.


\section{Small but finite flow inertia: \texorpdfstring{$\hat{Re}=\mathcal{O}(1)$}{hatRe=O(1)}}
\label{sec:nonzero-Re}

A feature of soft hydraulics problems is that the unidirectional flow solutions are derived under the lubrication approximation. As such, these solutions are \emph{approximate} solutions and, thus, are \emph{not} valid for arbitrary Reynolds number, unlike classical unidirectional flow solutions in ducts \cite{W06_book}. Specifically, when the reduced Reynolds number, $\hat{Re}$, is no longer vanishingly small, the inertial terms in equation \eqref{COLM-Z-O1} are no longer negligible. However, equations \eqref{COLM-X-O1} and \eqref{COLM-Y-O1} dictate that the pressure at each cross-section is still uniform at the leading order (in $\epsilon$), hence we can still construct a 1D model relating the pressure $P(Z)$ to the flow rate $Q$. 

Towards this end, as before, we can either introduce $H_e(Z)$, based on enforcing a Poiseuille-like law~\eqref{dP-Q-Heff}, or introduce the averaged channel height $\bar{H}(Z)$  as an approximation to $H_e(Z)$ in the same relation. As shown in section~\ref{sec:zero-Re} for $\hat{Re}\to0$, using $\bar{H}$ in place of $H_e$ commits a controllable error, and both approaches lead to similar results (as long as the deformation gradient is small). Instead of treating both cases for $\hat{Re}=\mathcal{O}(1)$, we refer the reader to the work by \citet{WC20}, who implemented the calculation based on $H_e(Z)$. In this section, we construct a 1D model using $\bar{H}(Z)$.

\subsection{Pressure distribution using an averaged deformed channel height}
\label{sec:R1-P}

To accomplish this task, the von K\'arm\'an--Pohlhausen approximation (see \S4-6.5 of White's book \cite{W06_book}) is employed to enforce a shape of the streamwise velocity profile, $V_Z^{2D}$, so that the flow rate in the deformed fluid domain can be obtained \cite{SWJ09,PPP15,IWC20}. That is, we assume a dimensionless parabolic axial velocity profile $V_Z^{2D}$, which is related to the dimensionless volumetric flow rate $Q$, as
\begin{equation}\label{Vz-Q-2D}
    V_Z^{2D}(Y,Z) = \frac{6QY[\bar{H}(Z)-Y]}{\bar{H}^3(Z)}.
\end{equation}
As discussed in the section~\ref{sec:zero-Re}, a profile as in equation~\eqref{Vz-Q-2D} is dictated by the Navier--Stokes equations for $\hat{Re}\to0$ (lubrication flow), and is generally valid for laminar flows \cite{W06_book}. An implicit assumption for using the velocity profile~\eqref{Vz-Q-2D} for finite $\hat{Re}$ is that flow inertia is weak: streamlines remain parallel and no recirculation occurs. Of course, this means that the theory developed herein is not valid in regimes in which transitional or turbulent flows occur. Indeed, the target application of our study is microfluidics, in which turbulent flows are not expected (or, generally possible) \citep{SSA04,B08}, although laminar flow with $\hat{Re}=\mathcal{O}(1)$ can be achieved \citep{SD19,DITT07,Lim14}.

Substituting equation \eqref{Vz-Q-2D} into equation \eqref{COLM-Z-O1} and integrating over $Y\in[0,\bar{H}(Z)]$, we obtain
\begin{equation}\label{Q-P-Re-2D}
    \frac{6}{5}\hat{Re}\frac{\rd}{\rd Z}\left[\frac{Q^2}{\bar{H}(Z)}\right]=-\frac{\rd P}{\rd Z}\bar{H}(Z)-\frac{12Q}{\bar{H}^2(Z)}.
\end{equation}
Observe that this expression, based on an equivalent 2D flow with $\bar{H}$ as the effective channel height, does not depend (or require integration) over $X$. 
{It should be noted that in the thin films literature \cite{RQM00,RACT10} inertial corrections to lubrication theory are also formulated, going to higher orders. Instead of assuming a parabolic velocity profile as in equation \eqref{Q-P-Re-2D}, a polynomial is used, and the coefficients are determined by incorporating the cross-sectional momentum equations, with their relevant boundary conditions, as well as the necessary corrections to the pressure. This approach is beyond the scope of the present work, however, as we consider wide channels ($\delta\ll1$), and there is no cross-sectional flow components ($V_X$ or $V_Y$) at the leading order \cite{CCSS17} in $\delta$ and $\epsilon$.}

Next, substituting $\bar{H}$ from equation $\eqref{Havg2}$ into equation \eqref{Q-P-Re-2D}, we once again obtain a separable first-order ODE for $P(Z)$. Imposing the outlet BC, $P(1)=0$, equation \eqref{Q-P-Re-2D} integrates to
\begin{multline}\label{P-Re-2D}
    P(Z)+\frac{3}{2}\xi P^2(Z)+\xi^2P^3(Z)+\frac{1}{4}\xi^3 P^4(Z) \\ 
    -\frac{6}{5}\hat{Re}\xi Q^2 P(Z)=12Q(1-Z),
\end{multline}
where $\xi= \lambda\mathcal{I}_1$ as above. As before, in the flow-controlled regime, $Q=1$, and $\Delta P$ is found implicitly from equation~\eqref{P-Re-2D}. Meanwhile, in the the pressure-controlled regime, after enforcing $P(0)=1$, equation~\eqref{P-Re-2D} becomes a quadratic in $Q$, and it has only one positive root:
\begin{equation}
    Q =\sqrt{\frac{25}{\hat{Re}^2\xi^2}+\frac{5}{24\hat{Re}\xi}\left(4+6\xi+4\xi^2+\xi^3\right)}-\frac{5}{\hat{Re}\xi} .
\end{equation}
This expression generalizes equation~\eqref{eq:Q_Re0} and shows the dependence on $\hat{Re}$ explicitly in the  inertial flow.

Since equation \eqref{P-Re-2D} is a polynomial in $P$, we can invert it to find the pressure distribution in the duct. Importantly, we expect $\rd P/\rd Z<0$ strictly for all $X\in[0,1]$ because of the assumption of laminar flow. Since $P(1)=0$, then $P(Z)>0$ for all $Z\in[0,1)$, which actually imposes an upper bound on the allowed values of $\hat{Re}$ and $\lambda$. To prove this bound, the leading-order term of the left-hand side of equation \eqref{P-Re-2D} is calculated to be  $(1-6\hat{Re}\xi Q^2/5)P$, as $Z\to 1^-$, while the right-hand side is positive. To ensure $P(Z)>0$ as $Z\to 1^-$, we must require that
\begin{equation}\label{eq:Re_upper_bound}
    \hat{Re}\lambda <\frac{5}{6\mathcal{I}_1Q^2}.
\end{equation}
Note that $\mathcal{I}_1$ is set by the solution of the corresponding elasticity problem (recall tables~\ref{tab:table-I} and \ref{tab:table-II}). 

\subsection{An extension and regularization via weak tension}
\label{sec:R1_reg}

At first glance, the restriction~\eqref{eq:Re_upper_bound} might be puzzling, but it actually ensures a continuous, and thus physical,  pressure distribution and wall deformation at the leading order. Since the local deformed height is linearly proportional to the local pressure at the leading order, prominent local deformation can be expected for sufficiently inertial flows and/or sufficiently compliant ducts. In the case for which the restriction~\eqref{eq:Re_upper_bound} is violated, the local deformation can be so large that it cannot transition smoothly near to zero at the outlet (to satisfy the boundary condition $P(1)= 0$, equivalently $\bar{H}(1)=1$). Thus, the solution~\eqref{P-Re-2D} breaks down for $\hat{Re}$ values that violate the restriction~\eqref{eq:Re_upper_bound}. 

In deriving equation~\eqref{P-Re-2D}, we used equation \eqref{Havg2}, which is a leading-order solution (in $\epsilon$) based on a plane strain configuration of the elastic wall's deformation field. This solution does not take into account the reaction forces imposed by connectors at the inlet and outlet of the duct. In this sense, we can think of the solid mechanics problem as being essentially a boundary layer problem. The Winkler-foundation-like mechanism (equation \eqref{Havg2}) is dominant outside the boundary layers, while some other mechanism plays a role within thin (boundary) layers near $Z=0,1$ to regularize the problem and account for the fact that the displacements in the vicinity of the inlet (or outlet) of the channel are usually restricted by external connections.

Since the bulging of the top wall unavoidably introduces stretching along $Z$ in the solid, a simple extension of equation \eqref{Havg2}, which can circumvent the restriction \eqref{eq:Re_upper_bound}, can be achieved by introducing weak constant tension into the formulation \cite{LP96}. Note the tension has to be ``weak'' to ensure the dominance of the Winkler-foundation-like mechanism. Other regularization mechanisms are also possible. For example, in the setting of elastic structures on top of thin fluid films, \citet{PL20} considered bending and gravity in addition to tension as regularization mechanisms. However, weak tension is arguably the simplest mechanism relevant to microchannels.  

Then, we may write down a governing equation for the deformed channel height \footnote{Note that equation \eqref{Havg-ext} would typically be written for the deformation, not the channel height. But, in our nondimensionalization, the deformation is simply $\bar{H}-1$.}:
\begin{equation}\label{Havg-ext}
    -\theta^2\frac{\rd^2 \bar{H}}{\rd Z^2} + \bar{H} -1 = \xi P.
\end{equation}
As motivated above, the dimensionless tension parameter $\theta^2\ll 1$. In this way, equation \eqref{Havg2} is precisely the outer solution of equation \eqref{Havg-ext} with $\theta^2=0$. { To give a physical expression for $\theta^2$, we transform equation \eqref{Havg-ext} back into dimensional form:
\begin{equation}
    f_t\frac{\rd^2 \bar{h}}{\rd z^2} + \mathcal{K}(\bar{h} - h_0) = p(z),
\end{equation}
where $f_t$ denotes the constant tension force per unit width (\si{\newton\per\meter}), and $\mathcal{K}= \mathcal{P}_c/(\xi h_0)$ is the \emph{effective} stiffness of the interface (\si{\pascal\per\meter}). Then, clearly, $\theta^2 = \xi f_t h_0/(\mathcal{P}_c \ell^2)$. 

The tension $f_t$ can arise from two physical scenarios. First, $f_t$ can arise due to stretching along $z$, which is caused by the bulging of the fluid--solid interface. In this scenario, $f_t$ can be estimated by averaging the elongation of the fluid--solid interface along $z$ \cite{HBDB14}:
\begin{equation}
f_t = \frac{E t^{\star}}{\ell}\int_0^\ell \frac{1}{2}\left(\frac{\rd \bar{h}}{\rd z}\right)^2  \,\rd z.
\end{equation}
Here, $t^{\star}$ represents the \emph{effective} thickness of the fluid--solid interface. 
If the wall is thin, we can take $t^{\star} = t$. However, if the compliant wall is thick, the displacement decays away from the fluid--solid interface, as shown for the thick-walled case in Ref.~\onlinecite{WC19}. In this case, taking $t^{\star} = t$ tends to overestimate the tension effect. Further considerations would be needed to estimate $t^{\star}$ in this case, which is beyond the scope of the current work.

In the second scenario, $f_t$ is provided by the pre-tension arising from external connectors. On the one hand, the pre-tension needs to be large enough so that the deformation induced stretch is negligible. On the other hand, the pre-tension needs to be small to ensure that $\theta^2\ll 1$. For the purposes of this paper, we focus on the effect of weak tension, which is consistent with our use of linear elasticity.} 

Next, taking $\rd/\rd Z$ of both sides of equation \eqref{Havg-ext} and substituting into equation \eqref{Q-P-Re-2D}, we obtain a nonlinear ODE for $\bar{H}(Z)$:
\begin{equation}\label{Q-P-Re-ext}
    \frac{3}{5}\hat{Re}\frac{\rd}{\rd Z}\left(\frac{Q^2}{\bar{H}^2}\right) = \frac{1}{\xi}\left(\theta^2\frac{\rd^3 \bar{H}}{\rd Z^3}-\frac{\rd \bar{H}}{\rd Z}\right)-\frac{12Q}{\bar{H}^3}.
\end{equation}

At the inlet and outlet, the top wall is { restricted from moving}, so the BCs for equation~\eqref{Q-P-Re-ext} are
    \begin{align}
    \bar{H}(0) = \bar{H}(1) &= 1, \label{eq:bc_h}\\
    \left.\frac{\rd^2 \bar{H}}{\rd Z^2}\right|_{Z=1} &= 0,\label{eq:bc_p_h}
\end{align}
where the BC \eqref{eq:bc_p_h} is a restatement of the outlet BC $P(1)=0$ in terms of $\bar{H}$ using equations~\eqref{Havg-ext} and \eqref{eq:bc_h}. Equations \eqref{Q-P-Re-ext}, \eqref{eq:bc_h} and \eqref{eq:bc_p_h} constitute a nonlinear \emph{two-point boundary-value problem}\citep{Keller76}. As before, in the flow-controlled situation, $Q=1$ and equation \eqref{Q-P-Re-ext} can be solved for $\bar{H}(Z)$ subject to the BCs \eqref{eq:bc_h}--\eqref{eq:bc_p_h}. In the pressure-controlled situation, $Q$ is found as an eigenvalue after imposing $P(0)=1$ on equations~\eqref{Q-P-Re-ext}, \eqref{eq:bc_h} and \eqref{eq:bc_p_h}.

Now, the restriction~\eqref{eq:Re_upper_bound} can be relaxed in the context of equation \eqref{Q-P-Re-ext}, in which the weak tension tends to restrain the wall deformation and, thus, regularizes the problem. Of course, the extent of regularization depends on the value of $\theta^2$. For example, if $\lambda=1.0$ and $\mathcal{I}_1=0.542754$ (for the thick-walled microchannel), then the upper bound of validity of the model is $\hat{Re} \approx 1.5$ for $\theta=0$. However, if $\theta^2 = 10^{-4}$, equation \eqref{Q-P-Re-ext} can be solved up to $\hat{Re}\approx 2.0$. If $\theta^2$ is further increased to $10^{-3}$, then equation \eqref{Q-P-Re-ext} can be solved up to $\hat{Re}\approx3.0$. {For such a large value of $\hat{Re}$, one can interpret the breakdown of equation \eqref{Q-P-Re-ext} as the breakdown of the lubrication theory and, potentially, as a sign that the ``full'' iNS equations need to be solved instead.} Next, we illustrate these observations and explain how equation \eqref{Q-P-Re-ext} was solved numerically.

\begin{figure*}[ht!]
    \centering
    \subfloat[]{\includegraphics[width=0.5\textwidth]{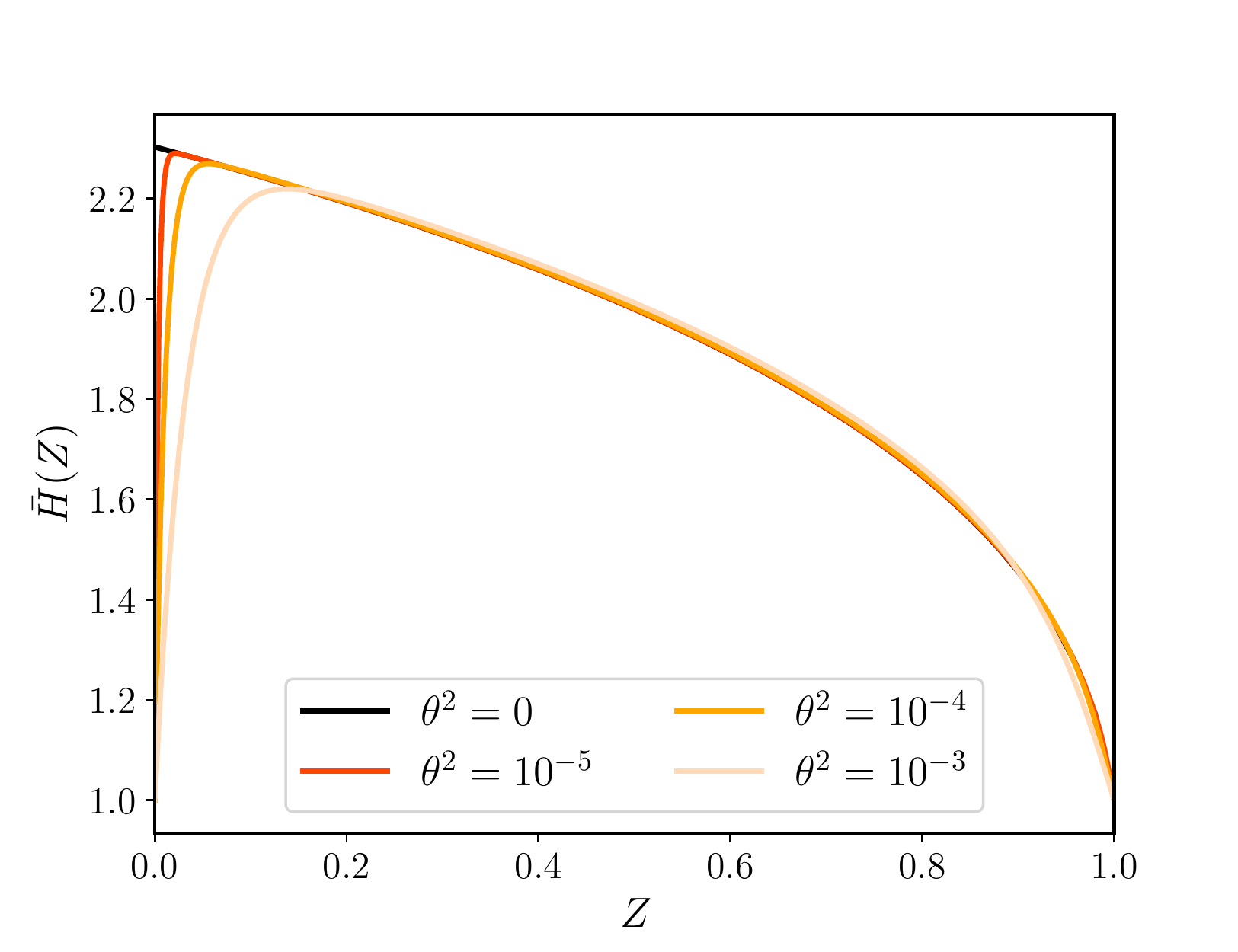}\label{fig:H-theta2}}\hfill
    \subfloat[]{\includegraphics[width=0.5\textwidth]{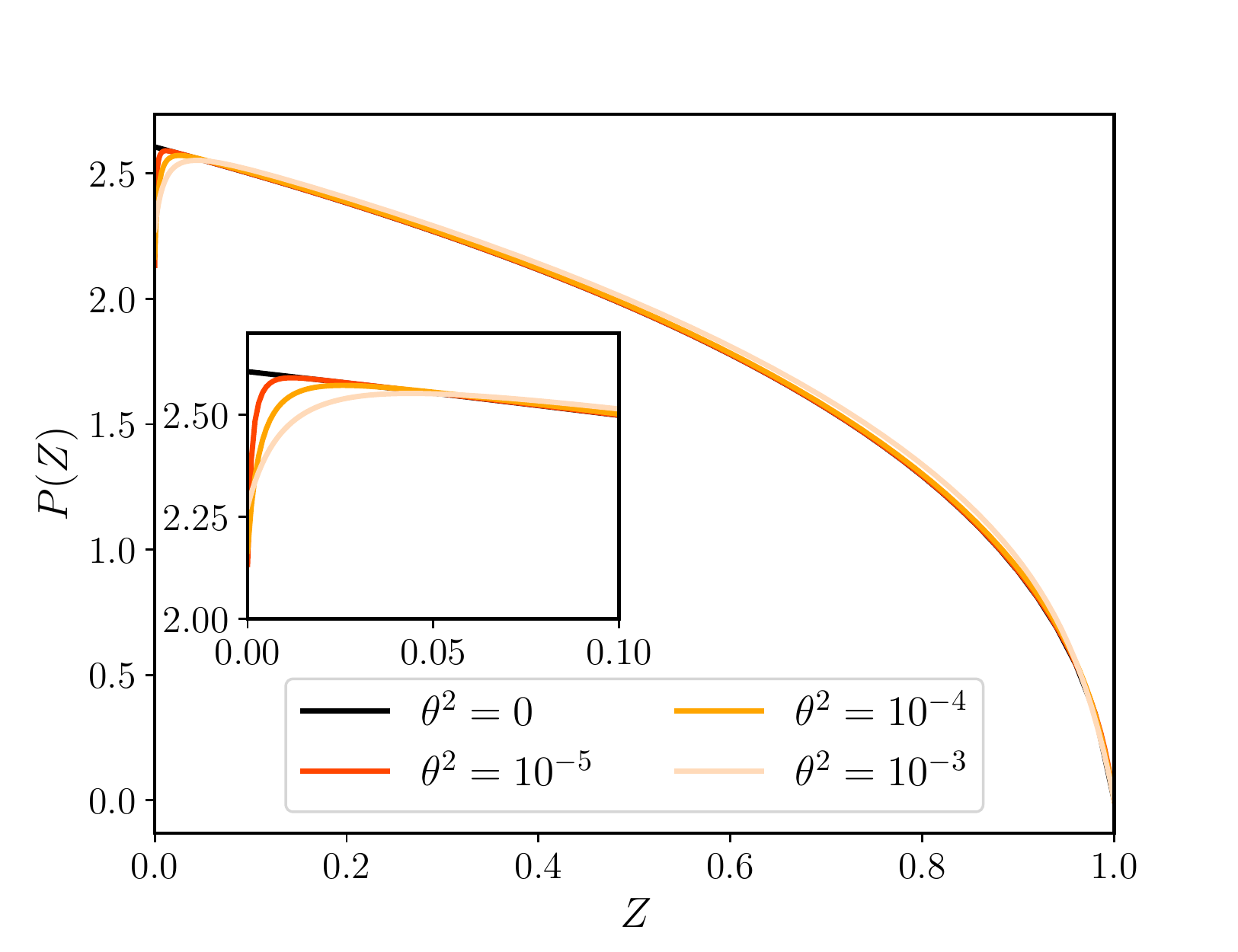}\label{fig:P-theta2}}
    \caption{(a) The deformed channel height $\bar{H}(Z)$ for different values of the tension parameter $\theta^2$. The { black} curves represent the outer solution of equation \eqref{Q-P-Re-2D}, while the {other (lighter)} curves are obtained using the ``full'' (numerical) solution of the two-point boundary-value problem consisting of equations \eqref{Q-P-Re-ext}, \eqref{eq:bc_h} and \eqref{eq:bc_p_h}. (b) The corresponding pressure distribution $P(Z)$. The { black} curves are obtained by substituting the solution of equation \eqref{Q-P-Re-2D} into equation \eqref{Havg-ext}, while the {other (lighter)} curves are similarly obtained from ``full'' (numerical) solution of equations \eqref{Q-P-Re-ext}, \eqref{eq:bc_h} and \eqref{eq:bc_h}. In both panels, we fixed $Q=1$, $\hat{Re}=1.0$, and $\xi=0.5$.}
\end{figure*}

\begin{figure*}[ht!]
    \centering
    \subfloat[]{\includegraphics[width=0.5\textwidth]{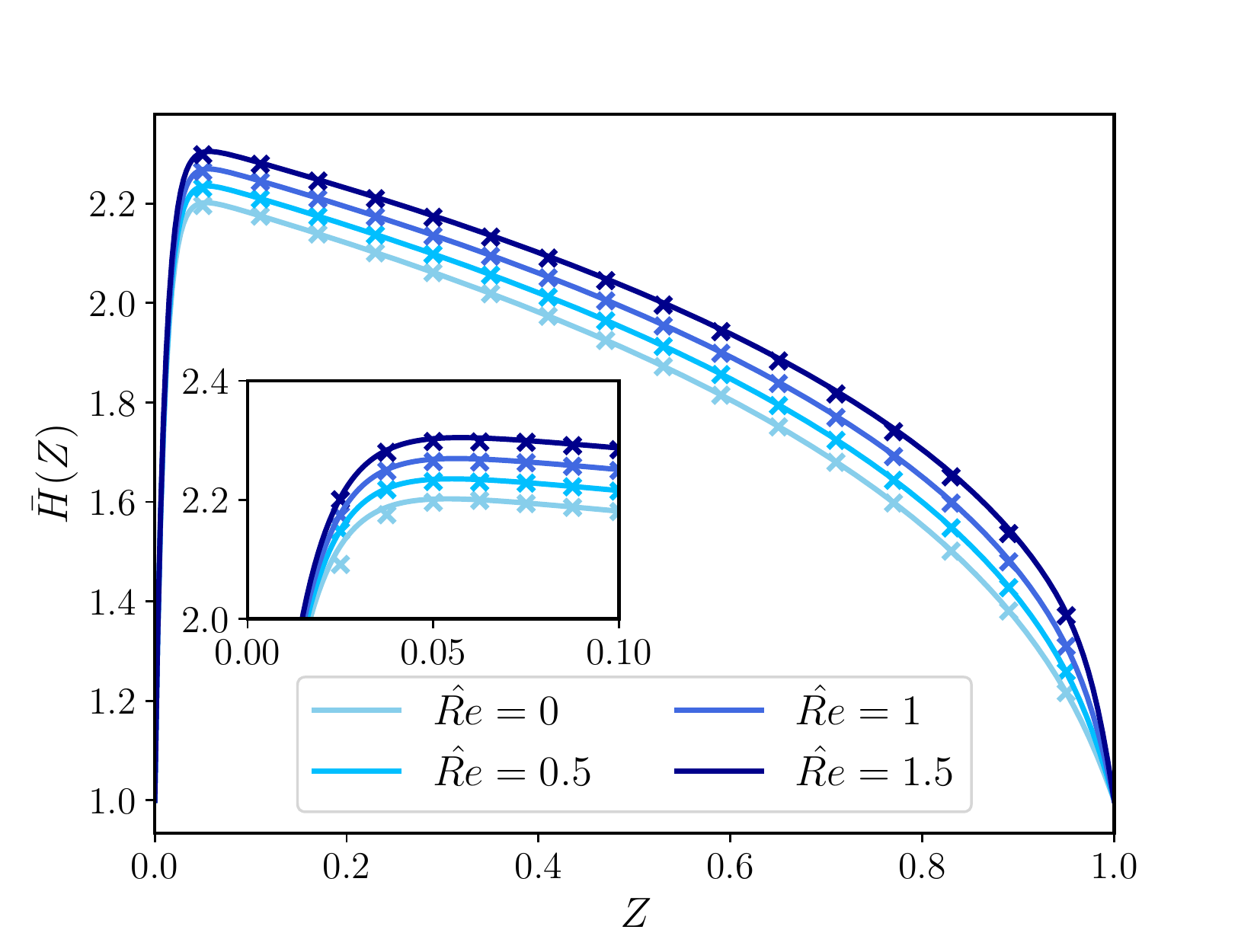}\label{fig:H-Re}}\hfill
    \subfloat[]{\includegraphics[width=0.5\textwidth]{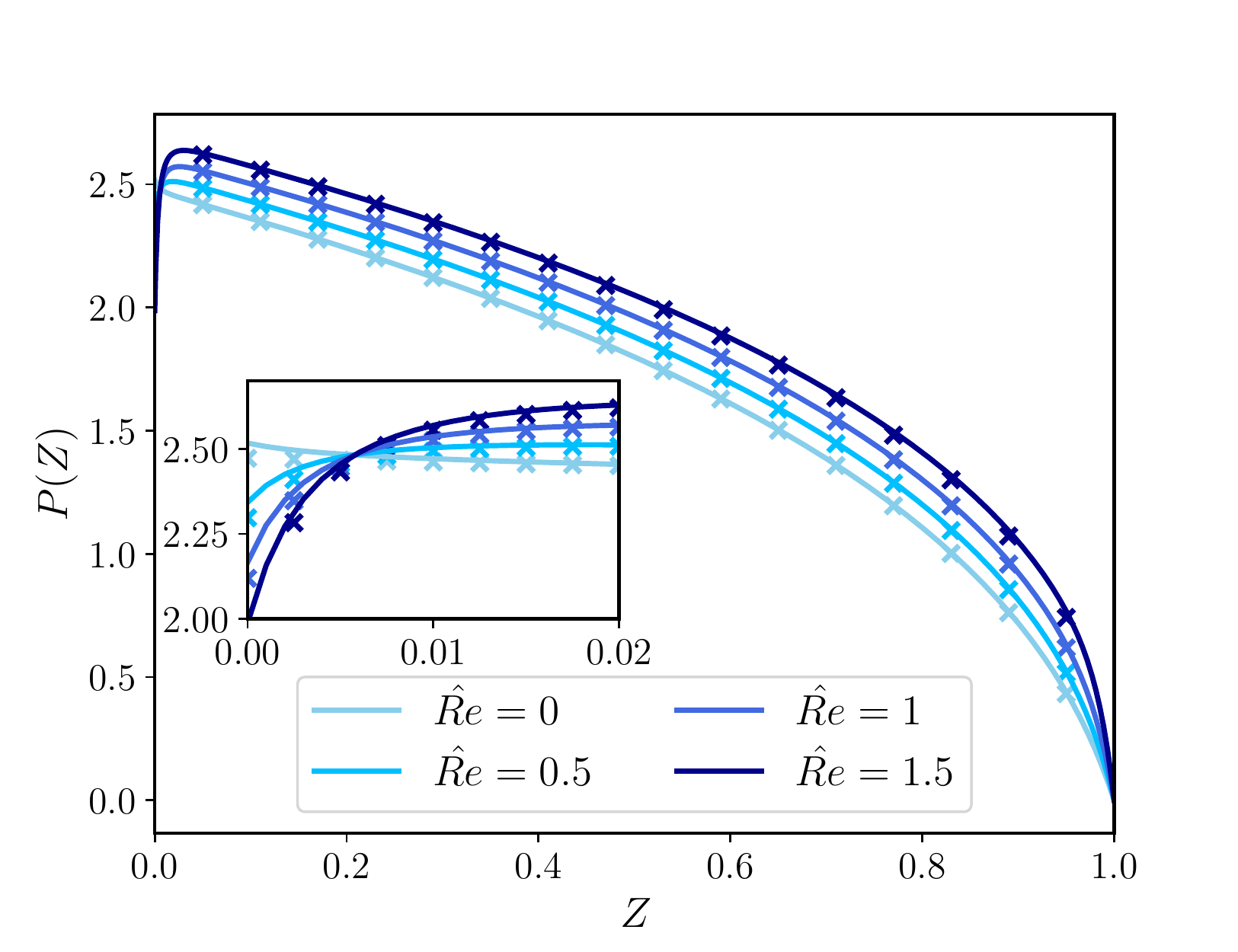}\label{fig:P-Re}}
    \caption{(a) The deformed channel height $\bar{H}(Z)$ for different values of the reduced Reynolds number $\hat{Re}$. The solid curves represent the numerical solution of equation \eqref{Q-P-Re-ext}, while the { symbols represent} the asymptotic solution (see equations \eqref{Hasym} and  \eqref{Hasym0Re}). (b) The corresponding pressure distributions $P(Z)$. The solid curves are obtained by substituting the solution of equation \eqref{Q-P-Re-ext} into equation \eqref{Havg-ext}, while the { symbols} are the asymptotic solution (see equations \eqref{Pasym} and \eqref{Hasym0Re}). The agreement between the asymptotic and numerical solutions is so good that the curves mostly overlap. In both panels, we  fixed $Q=1$, $\xi=0.5$, and $\theta^2=10^{-4}$.}
\end{figure*}

\subsection{Illustrated examples}

Depending on the top wall's geometry, $\xi$ in equations \eqref{Q-P-Re-2D} and \eqref{Q-P-Re-ext} will take different forms, such as the thick wall case and plate-like top wall case introduced in sections~\ref{sec:Re0_thick} and \ref{sec:Re0_plate}, respectively. To make our discussion general, instead of considering the two cases separately, we regard $\xi$ and $\theta$ as characteristic system parameters and discuss the corresponding solutions of equations \eqref{Q-P-Re-2D} and \eqref{Q-P-Re-ext} to illustrate the regularization introduced in section~\ref{sec:R1_reg}. Equation \eqref{Q-P-Re-2D} can be solved in two steps. First, invert equation \eqref{P-Re-2D} to get $P(Z)$. Second, substitute $P(Z)$ into equation \eqref{Havg2} to get $\bar{H}$. As for equation \eqref{Q-P-Re-ext}, we use the {\tt solve\_bvp} routine from the SciPy stack \citep{SciPy} to obtain a numerical solution of the nonlinear two-point boundary value problem. After obtaining $\bar{H}(Z)$, equation \eqref{Havg-ext} can be used to obtain $P(Z)$.

First, we investigate the tension effect by varying $\theta^2$ in equation \eqref{Q-P-Re-2D} while keeping $Q$, $\hat{Re}$ and $\xi$ fixed. As shown in figure~\subref*{fig:H-theta2}, with $\theta^2\ll 1$, the solutions to equations \eqref{Q-P-Re-2D} and \eqref{Q-P-Re-ext} do not differ much from each other along most of the domain $Z\in[0,1]$, as required by the dominance of the Winkler-foundation-like mechanism of deformation. Since equation \eqref{Q-P-Re-2D} only satisfies $\bar{H}(1)=1$, in principle, two boundary layers could be expected near $Z=0$ and $Z=1$, respectively, to fulfill the remaining boundary conditions from equations \eqref{eq:bc_h}--\eqref{eq:bc_p_h}. However, as we have discussed in section~\ref{sec:R1-P}, with $\theta^2=0$, equation \eqref{P-Re-2D} indicates that, $P$ varies linearly with $Z$ as $Z\to 1^-$. Since $\bar{H}$ is linearly proportional to $P$ at the leading order in $\theta$, $\bar{H}$ should be linear in $Z$ as $Z\to 1^-$, hence $\rd^2 \bar{H}/\rd Z^2 \to 0$ as $Z\to 1^-$. In other words, the outer solution actually satisfies the boundary condition \eqref{eq:bc_p_h}. Therefore, there is no boundary layer located near $Z=1$. This fact can also be seen in  figure~\subref*{fig:H-theta2}, where the left boundary layer is prominent (becoming thicker as $\theta^2$ is increased), while the outer solution agrees well with the full numerical solution near $Z=1$ for all values of $\theta^2$ shown.

The effect of $\theta^2$ on $P(Z)$ is shown in figure~\subref*{fig:P-theta2}. The key takeaway from this plot is that, while equation \eqref{Q-P-Re-2D} always predicts $P(Z)$ to be a decreasing function of $Z$, a positive pressure gradient is observed near $Z=0$ in the numerical solution to equation~\eqref{Q-P-Re-ext} for all  $\theta^2\ne0$ considered. The reason for this positive pressure gradient near the inlet is that, due to the restriction on the displacement at $Z=0$, the area of the cross-section undergoes a sharp change near $Z=0$. Since the flow rate is fixed at steady state, the axial velocity has to quickly reduce near $Z=0$. The observed positive pressure gradient  facilitates this deceleration of the flow.

Next, we address the effect of fluid inertia by varying $\hat{Re}$. In this case, we fix $Q=1$, $\xi=0.5$ and $\theta^2=10^{-4}$. As shown in figure~\subref*{fig:H-Re}, as $\hat{Re}$ increases, larger deformation of the wall is observed. Also, the deformation gradient along $Z$ is larger for higher $\hat{Re}$ because the pressure displays sharper decrease with the increase of $\hat{Re}$, which can be clearly seen in figure~\subref*{fig:P-Re}. Notably, $\rd P/\rd Z>0$ is also observed for the three cases of $\hat{Re}\neq 0$, which can be explained as before. However, $\rd P/\rd Z$ remains negative in the case of $\hat{Re}=0$. This is because, in this case of negligible fluid inertia, the deceleration of the flow near the inlet is not as large as the other cases, thus the positive pressure gradient is not necessary. Finally, we mention that instead of solving equation \eqref{Q-P-Re-ext} numerically, we are able to obtain a uniformly valid asymptotic solution for $\bar{H}(Z)$ and $P(Z)$ using the method of matched asymptotic expansions \citep{H13}. In particular, for the special case of $\hat{Re}=0$, we are able to obtain explicit formulae for both $\bar{H}(Z)$ and $P(Z)$. The details of this calculation are provided in appendix~\ref{sec:app}. The dashed curves in figures~\subref*{fig:H-Re} and \subref*{fig:P-Re} demonstrate that these asymptotic solution (equations \eqref{Hasym} and \eqref{Pasym} in appendix~\ref{sec:app}) agrees well with the numerical solution.

\begin{table*}
\caption{Typical values of the dimensional and dimensionless parameters arising from equation \eqref{Havg-ext}.}
\begin{tabular}{l@{\qquad} l@{\qquad} l@{\qquad} l}
\toprule
Name & Variable & Typical value & Unit \\
\hline
 channel's length & $\ell$ & $1.0$ & \si{\centi\meter} \\
channel's undeformed height & $h_{0}$ &$25$ & \si{\micro\meter} \\
channel's width & $ w $ & $500$ & \si{\micro\meter}\\
top wall's thickness & $t$ & $2.0$ & \si{\milli\meter}\\
solid's Young's modulus & $E$ &$1.5$ & \si{\mega\pascal}  \\
solid's Poisson's ratio & $\nu$ & $0.5$ & -- \\
fluid's dynamic viscosity & $\mu$ & $1.0\times10^{-3}$ & \si{\pascal\second} \\
fluid's density & $\rho$ &$1.0\times10^{3}$ & \si{\kilo\gram\per\meter\tothe{3}} \\
inlet flow rate & $q$ & See table~\ref{tab:response-q} & \si{\micro\litre\per\minute} \\
tension force per unit width & $f_t$ & 400 & \si{\newton\per\meter}\\
characteristic velocity scale & $ \mathcal{V}_c = {q}/(wh_0) $  & -- & \si{\meter\per\second}\\
characteristic pressure scale & $\mathcal{P}_c ={\mu \mathcal{V}_c}/{(\epsilon h_0)}$ & -- & \si{\kilo\pascal} \\
pressure drop & $\Delta p = p(z=0)$ & See table~\ref{tab:response-q} & \si{\kilo\pascal}\\
maximum pressure & $p_{\max}=\displaystyle \max_{ 0\leq z\leq \ell} p(z)$ & See table~\ref{tab:response-q} & \si{\kilo\pascal} \\
maximum channel's deformed height & $\bar{h}_{\max}$ = $\displaystyle \max_{0\leq z\leq \ell} \bar{h}(z)$ & See table~\ref{tab:response-q} & \si{\micro\meter} \\
\hline
channel's height-to-length aspect ratio & $\epsilon={h_0}/{\ell}$ & $0.0025$ &  --\\
channel's height-to-width aspect ratio & $\delta = {h_0}/{w} $ & $0.05$ & --\\
reduced Reynolds number & $\hat{Re}={\epsilon\rho q}/{(w\mu)}$  & See table~\ref{tab:response-q} &  --\\
dimensionless spring stiffness & $\xi=\lambda \mathcal{I}_1$  ($\lambda = {w\mathcal{P}_c}/{(h_0\bar{E})},\ \mathcal{I}_1=0.542754$) &  See table~\ref{tab:response-q} &  --\\
tension coefficient & $\theta^2 = {f_t h_0 \xi}/{(\mathcal{P}_c \ell^2)}= {f_t w \mathcal{I}_1}{(\bar{E} \ell^2)} $ & $5.427\times 10^{-4}$ & -- \\
\botrule
\end{tabular}
\label{table:param}
\end{table*}

\begin{table}[ht]
\caption{Calculated steady-state responses of the microchannel system under different flow rate with the parameters specified in table~\ref{table:param}.}
    \begin{tabular}{l@{\quad} l@{\quad} l@{\quad} l@{\quad} l@{\quad} l }
    \toprule
     $q$  & $\hat{Re}$ & $\xi$ & $\Delta p$ & $p_{\max}$ & $\bar{h}_{\max}$   \\
     (\si{\micro\litre\per\minute})  & (--) & (--) & (\si{\kilo\pascal}) & (\si{\kilo\pascal}) & (\si{\micro\meter}) \\
     \hline
     $1500$ & $0.125$ & $0.1737$ & $140.96$ & $140.96$ & $42.55$ \\
     $6000$ & $0.5$ & $0.6947$ & $250.55$ & $266.16$ & $59.74$ \\
     $12000$ & $1.0$ & $1.3895$ & $258.27$ & $366.90$ & $73.61$ \\
    \botrule
    \end{tabular}
    \label{tab:response-q}
\end{table}

{ As a supplement to our discussion above, typical values of the dimensional and dimensionless variables of a microchannel with a thick top wall are summarized in table~\ref{table:param}. Here $t^2/w^2=16\gg 1$, thus equation \eqref{lam-thick} and table~\ref{tab:table-I} from section~\ref{sec:Re0_thick} are applicable. The steady responses of the system under different flow rates are calculated from equation \eqref{Havg-ext} and tabulated in table~\ref{tab:response-q}. With the increase of the flow rate, the pressure drop, the maximum pressure within the channel, and the maximum deformation of the interface are increasing. As we have discussed, when the flow inertia is small (smaller $\hat{Re}$), the maximum pressure occurs at the inlet of the channel. However, if the flow inertia is prominent, there is a positive pressure gradient near the inlet and thus, the maximum pressure is ``pushed'' inwards, away from the inlet.}


\section{Conclusion}
\label{sec:conclusion}

In the spirit of Frank M.\ White's summary of unidirectional flows \cite{W06_book} in non-circular ducts, we critically discussed weakly-unidirectional flows (under a lubrication scaling) in compliant ducts of initially rectangular cross-section, for both the vanishing and the finite Reynolds number cases. In doing so, we contributed to the recently developed theory of \emph{soft hydraulics}. Attention was paid to the hydraulic resistance of such conduits during steady viscous flow (\textit{i.e.}, the flow rate--pressure drop relations, which are now nonlinear). In particular, we derived 1D reduced models from 3D results on fluid--structure interaction. In doing so, we synthesized and unified a variety of previous models (some justified only by empirical considerations). This kind of reduction has been sought (and is of general interest \citep{PPP15}) for practical design considerations of microfluidic systems \citep{MY18,MY19,S17_LH,P21}, such as for calibrating optics-free non-contact measurement techniques \citep{Dhong18}.

For inertialess unidirectional flow in a compliant duct, the pressure varies nonlinearly along the streamwise direction due to the FSI between the viscous fluid flow and the compliant wall. Due to the slenderness and shallowness of the duct, we are able to relate the nonlinear pressure gradient $\rd p/\rd z$ to the flow rate $q$ at steady state. By introducing the concept of an effective channel height, we recovered the form of the classical Poiseuille-like law and, at the same time, reduced the original 3D flow problem to an equivalent 2D one. 

Although averaged deformed channel heights have been used in the literature, the validity of such models was not previously established. We found that the averaged channel height~\eqref{havg} can be a good approximation to the consistent effective height introduced in equation~\eqref{heff}. This conclusion is important because the averaged-height models yield explicit flow rate--pressure drop relations, and are easily compared to other geometries such as axisymmetric cases. Interestingly, we showed that the averaged channel height has a universal expression as $\Bar{H}(Z) = 1+\xi P(Z)$, where $\xi=\lambda\mathcal{I}_1$, for \emph{both} thick-walled and thinner, plate-like-walled top walls. Even though the formula for the dimensionless compliance coefficient $\xi$ is different in the two cases, we have justified the observation (from the end of section~\ref{sec:heff}) that a wide and shallow microchannel's top wall behaves like a Winkler foundation \cite{W67,DMKBF18}, in which the averaged channel height is determined by the local pressure and a proportionality constant.

The reduction of the 3D FSI problem to a 1D model using the averaged height concept also allowed us to generalize the textbook concept of a friction factor \citep{W06_book,P11} to compliant ducts. We showed that the soft hydraulic system's Poiseuille number $Po$ (product of the Fanning friction factor $C_f$ and the Reynolds number) can be between $1$ and $4$ times larger than that for a rigid duct. Importantly, for the compliant duct, both $C_f$ and $Po$ depend on the streamwise coordinate due to the non-constant pressure gradient. This novel result extends the laminar portion of the Moody diagram, in which roughness is unimportant, via a new compliance parameter that is important in microfluidics.

Additionally, we showed how to incorporate weak but finite flow inertia in the previous $Re\to0$ models. The finite-$Re$ model breaks down beyond a certain value of the product of $Re$ and a compliance parameter $\lambda$. Weak tension near the inlet and outlet of the reduced 1D model was introduced to regularize this breakdown and to obtain uniformly valid pressure distributions (in the sense of matched asymptotics).

The present results pave the way towards understanding more complex unsteady soft hydraulic phenomena. Specifically, with all this in mind, we would like to revisit and extend the linear stability results from \cite{WC20} to the reduced-order models derived herein. This analysis could shed new insight on ``ultrafast mixing'' and multifold reduction of the critical Reynolds number recently observed in experiments on flow in compliant microchannels \cite{VK13,NS15,KB16}. Elastic walls (or coatings) have been shown to alter the turbulent boundary layer energy budget in channels \cite{G02,RB20}, thus the transition to turbulence in soft hydraulic systems \cite{KB16} is also expected to have nontrivial departures from the classical picture.

\section*{Acknowledgements}
This paper is dedicated, with respect and admiration, to Prof.\ Frank M.\ White on the occasion of his 88\textsuperscript{th} anniversary.

I.C.C.\ thanks the Department of Mechanical Engineering at IIT Kharagpur for its hospitality during his visit there in December 2019, during which the idea for this work was conceived. Insightful discussions and input from J.\ Chakraborty and P.\ Karan, on Winkler foundations and 2D models, and from V.\ Anand and K.A.\ Flack, on friction factors in pipes, are kindly acknowledged. 

I.C.C.'s visit to IIT Kharagpur and this research was enabled by the Scheme for Promotion of Academic and Research Collaboration (SPARC), a Government of India Initiative, under Project Code SPARC/2018-2019/P947/SL. Additionally, X.W.\ and I.C.C.\ were partially supported by the US National Science Foundation under grant No.\ CBET-1705637.

\section*{Data availability statement}
Data sharing is not applicable to this article as no new data were created or analyzed in this theoretical study. Python script files for generating the plots, which are based on the equations in the text, are openly available in the Purdue University Research Repository at \href{https://dx.doi.org/10.4231/37PY-K896}{doi:10.4231/37PY-K896}, and/or upon reasonable request to the corresponding author.

\appendix
\renewcommand{\theequation}{A\arabic{equation}}
\section{Matched asymptotic solution for the 1D model with weak tension}
\label{sec:app}

For $\theta^2\ll 1$, equation \eqref{Q-P-Re-ext} subject to the BCs~\eqref{eq:bc_h}--\eqref{eq:bc_p_h} represents a singular perturbation problem \citep{H13}. The outer solution $\bar{H}_o(Z)$, which satisfies $\bar{H}_o(1)=1$, is found by setting $\theta^2=0$:
\begin{equation}\label{outer-sol}
    \frac{1}{\xi}\left[\frac{1}{4}(\bar{H}_o^4-1)-\frac{6}{5}\hat{Re}\xi Q^2({\bar{H}_o-1})\right]= 12Q(1-Z).
\end{equation}
Substituting equation \eqref{Havg2} into the above, we recover equation \eqref{P-Re-2D} as the outer solution for the pressure.

In the boundary layer near $Z=0$ (``left'' boundary layer), we introduce the rescaled coordinate $\hat{Z}=Z/\theta$. 
Denote the left inner solution as $\bar{H}_l(\hat{Z})$. Then, in terms of these new variables, equation \eqref{Q-P-Re-ext} is transformed into
\begin{equation}\label{left-eq1}
    \frac{3}{5}\hat{Re}\frac{\rd }{\rd\hat{Z}}\left(\frac{Q^2}{\bar{H}_l^2}\right) = \frac{1}{\xi}\left(\frac{\rd^3\bar{H}_l}{\rd \hat{Z}^3}-\frac{\rd \bar{H}_l}{\rd \hat{Z}}\right)+\theta\frac{12Q}{\bar{H}_l^3}.
\end{equation} 
At the leading order, the last term in equation~\eqref{left-eq1} is negligible, and we integrate once to obtain
\begin{equation}\label{left-eq2}
    \frac{3}{5}\hat{Re}\xi \frac{Q^2}{\bar{H}_l^2}=\frac{\rd ^2\bar{H}_l}{\rd \hat{Z}^2}-\bar{H}_l+C_1.
\end{equation}

Now, consider the behavior of  equation~\eqref{left-eq2} in the phase plane $(\mathscr{H},\mathscr{F})$, where we have defined $\mathscr{H}:=\bar{H}_l$ and $\mathscr{F}:=\rd \bar{H}_l/\rd \hat{Z}$; $\hat{Z}$ parametrizes integral curves (\textit{i.e.}, solutions) in this plane. Equation~\eqref{left-eq2} becomes
\begin{align}
    \frac{\rd \mathscr{H}}{\rd \hat{Z}} &= \mathscr{F}, \label{left-eq-pp-1}\\
    \frac{\rd \mathscr{F}}{\rd \hat{Z}} &= 
    \frac{3}{5}\hat{Re}\xi \frac{Q^2}{\mathscr{H}^2} + \mathscr{H} - C_1. \label{left-eq-pp-2}
\end{align}
Fixed points of the system~\eqref{left-eq-pp-1}--\eqref{left-eq-pp-2} are such that the right-hand sides vanish. Although the expression for the fixed point $(\mathscr{H}^\star,\mathscr{F}^\star)$ with and $\mathscr{H}^\star>0$ and $\mathscr{F}^\star=0$ is lengthy, it can be found. The solution of equation~\eqref{left-eq2} as $\hat{Z}\to\infty$ and $\rd \bar{H}_l/\rd \hat{Z}\to0$ should match the outer solution  $\bar{H}_o$ as ${Z\to0}$. Therefore, $\mathscr{H}^\star$ must be chosen to be precisely $\bar{H}_o(0)$, which is the positive real root of equation~\eqref{outer-sol} with $Z=0$. Consequently, without needing the explicit formula for $\mathscr{H}^\star$, we obtain: \begin{equation}
    C_1 = \frac{3}{5}\hat{Re}\xi \frac{Q^2}{{\bar{H}_o(0)^2}} + \bar{H}_o(0).
\end{equation}

Now, the inner solution in the left boundary layer is the integral curve in the $(\mathscr{H},\mathscr{F})$ plane starting at $\mathscr{H}=1$ and ending at $\mathscr{H}=\bar{H}_o(0)$. To construct this curve, multiply both sides of equation~\eqref{left-eq2} by $\rd \bar{H}_l/\rd \hat{Z}$, and obtain a first integral:
\begin{equation}\label{left-eq3}
    \left(\frac{\rd \mathscr{H}}{\rd \hat{Z}}\right)^2 = - \frac{6}{5}\hat{Re}\xi \frac{Q^2}{\mathscr{H}} + \mathscr{H}^2 - 2 C_1\mathscr{H} + C_2 .
\end{equation}
To ensure that $\mathscr{H}^\star=\bar{H}_o(0)$ remains the desired fixed point of the ODE, the constant of integration must be
\begin{equation}
    C_2 = \frac{12}{5}\hat{Re}\xi \frac{Q^2}{\bar{H}_o(0)} + \bar{H}_o(0)^2.
\end{equation}
Then, equation \eqref{left-eq3} can be rewritten as:
\begin{multline}\label{left-eq4}
    \left(\frac{\rd \mathscr{H}}{\rd \hat{Z}}\right)^2 = [\mathscr{H}-\bar{H}_o(0)]^2\left\{1- \frac{6}{5}\hat{Re}\xi\frac{Q^2}{\mathscr{H}\bar{H}_o(0)^2}\right\} .
\end{multline}

Equation \eqref{left-eq4} is  separable, so its solution can be written as
\begin{equation}\label{left-eq5}
    \int_{1}^{\bar{H}_l} \frac{\rd \mathscr{H}}{[\bar{H}_o(0)-\mathscr{H}] \sqrt{1- \frac{6}{5}\hat{Re}\xi\frac{Q^2}{\mathscr{H}\bar{H}_o(0)^2}}} =  \hat{Z},
\end{equation}
where positive root is taken because it is expected that $\rd \mathscr{H}/\rd \hat{Z}>0$ and thus, $\bar{H}_o(0)>\mathscr{H}$, in the boundary layer.
Performing the integration in equation \eqref{left-eq5} yields an implicit solution:
\begin{multline}\label{left-eq5b}
    -2\left[\tanh^{-1}\left(\sqrt{1-\frac{m}{\bar{H}_l}}\right)- \tanh^{-1} (\sqrt{1-m}) \right]\\
    +\frac{2}{\sqrt{1-\frac{m}{\bar{H}_o(0)}}} \left[
   \tanh^{-1}\left( \sqrt{\frac{1-\frac{m}{ \bar{H}_l}}{1-\frac{m}{\bar{H}_o(0)}}} \right) \right.\\
   -
   \left.
   \tanh^{-1}\left( \sqrt{\frac{1-m}{1-\frac{m}{\bar{H}_o(0)}}} \right) \right] = \hat{Z},
\end{multline}
where $m = 6\hat{Re}\xi Q^2/[5\bar{H}_o(0)^2]$. Observe that if the criterion in equation~\eqref{eq:Re_upper_bound} is satisfied then  $m<1$ follows, which is required for the solution \eqref{left-eq5} to exist. Therefore, the restriction~\eqref{eq:Re_upper_bound} is needed to obtain a meaningful outer solution to equation \eqref{outer-sol}. In the case for which the criterion~\eqref{eq:Re_upper_bound} is violated, this asymptotic analysis will break down, which suggests that tension is no longer a sufficiently small effect. In that case, we can solve equation~\eqref{Q-P-Re-ext} numerically. 

Inverting equation \eqref{left-eq5b} to get an explicit expression for $\bar{H}_l(\hat{Z})$ is nontrivial. However, for the special case of  $\hat{Re}= 0$, equation \eqref{left-eq5} immediately gives an explicit solution:
\begin{equation}\label{left-eq6b}
   \bar{H}_l(\hat{Z}) = \bar{H}_o(0) + [1-\bar{H}_o(0)]e^{-\hat{Z}} \qquad (\hat{Re}=0).
\end{equation}

As for the right boundary, near $Z=1$, the ODE does not exhibit a boundary layer structure for $\theta^2\to0$, as we discussed in section~\ref{sec:R1-P}. This fact is also shown by figure~\subref*{fig:H-theta2},  
from which it is evident that the numerical solutions of the ``full'' ODE agree well with the leading-order outer solution (outside the left boundary layer), for any $\theta^2 \ll 1$.

The composite solution is obtained after subtracting the common part between inner and outer solutions: 
\begin{multline}\label{Hasym}
    \bar{H}(Z) \sim \bar{H}_a (Z) = \bar{H}_l(Z/\theta) + \bar{H}_o(Z) - \bar{H}_o(0),\\ 
    (\theta^2 \ll 1)
\end{multline}
with $\bar{H}_l$ and $\bar{H}_o$ given (implicitly) by equations \eqref{left-eq5b} and \eqref{outer-sol}, respectively. Equation \eqref{Havg-ext} can be used to obtain the asymptotic solution for $P$. The leading-order terms are
\begin{multline}\label{Pasym}
    P(Z) \sim P_a(Z) = \frac{1}{\xi} \left(-\frac{\rd^2 \bar{H}_l}{\rd \hat{Z}^2} + \bar{H}_a -1\right) \\
    = \frac{1}{\xi}\Bigg\{ \bar{H}_o(Z) -1 -\frac{3}{5}\hat{Re}\xi Q^2 \left[ \frac{1}{\bar{H}_l(Z/\theta)^2} - \frac{1}{\bar{H}_o(0)^2}\right] \Bigg\},\\ 
    (\theta^2\ll 1 )
\end{multline}
where we have used equation \eqref{left-eq2} to compute $\rd^2\bar{H}_l/\rd \hat{Z}^2$. 

For $\hat{Re}= 0$, using equation~\eqref{left-eq6b}, the composite solution can be explicitly written as
\begin{multline}\label{Hasym0Re}
    \bar{H}(Z) \sim \bar{H}_a(Z) = \left[1-(1+48Q\xi)^{1/4}\right]e^{-Z/\theta} \\
    + \left[1+48Q\xi(1-Z)\right]^{1/4} \qquad (\theta^2\ll1,\;\hat{Re}=0).
\end{multline}
Substituting equation \eqref{Hasym0Re} into equation \eqref{Havg-ext} (or, setting $\hat{Re}=0$ in equation \eqref{Pasym}), we obtain the matched asymptotic solution for the pressure distribution as well:
\begin{multline}\label{Pasym0Re}
    P(Z) \sim P_a(Z) = \frac{1}{\xi}\left\{\left[1+48\xi Q(1-Z)\right]^{1/4}-1\right\}\\
    (\theta^2\ll1,\;\hat{Re}= 0).
\end{multline}

\section*{References}

\bibliography{references.bib}

\end{document}